\documentclass[
  aps,
  onecolumn,
  showkeys,
  amsmath,
  amssymb
]{revtex4-2}

\usepackage{bm}
\usepackage{mathtools}
\usepackage{physics}
\usepackage{dsfont}

\usepackage{graphicx}
\usepackage{dcolumn}
\usepackage{array}
\usepackage{multirow}
\usepackage{tabularx}
\usepackage{float}
\usepackage{overpic}

\usepackage[caption=false]{subfig}

\usepackage{orcidlink}
\usepackage{enumitem}
\usepackage{verbatim}
\usepackage{soul}
\usepackage{xcolor}

\definecolor{darkBlue}{rgb}{0.08,0.13,0.4}

\hypersetup{
    colorlinks=true,
    citecolor=blue,
    linkcolor=blue,
    urlcolor=blue
}

\begin{document}

\preprint{APS/123-QED}

\title{Perfect Single-Excitation Mirroring in Long-Range Spin Chains}

\author{C.C. Nelmes$^{1,2,*}$\orcidlink{0009-0002-1686-6282}, I. D'Amico$^{1,2}$\orcidlink{0000-0002-4794-1348} and T.P. Spiller$^{1,2}$\orcidlink{0000-0003-1083-2604}}

\affiliation{$^1$School of Physics, Engineering and Technology, University of York, York, YO10 5DD, United Kingdom}

\affiliation{$^2$York Centre for Quantum Technologies, University of York, York, YO10 5DD, United Kingdom}

\affiliation{$^*$Author to whom any correspondence should be addressed.}

\email{c.nelmes@york.ac.uk, irene.damico@york.ac.uk, timothy.spiller@york.ac.uk}


\begin{abstract}
\noindent Long-range interacting systems are naturally present within quantum and classical systems alike. In quantum information processing, communication channels between distant processing units can themselves participate in computation via gate operations, entanglement generation, and/or quantum state transfer, making them fundamentally more versatile than classical wires. Ideal candidates for such channels are systems that allow perfect single-excitation mirroring (PSEM) or total mirror inversion of any single-excitation state, particularly via linear spectral spacing. However, very few examples of PSEM-solutions due to a completely linear spectrum for moderate system sizes $N \sim10$, are known to date. In this work, utilising an iterative inverse eigenvalue procedure, we construct PSEM protocols in both power-law long-range and next-nearest-neighbour interacting systems, which satisfy mirror symmetry and complete spectral linearity. We present the necessary site spacings and on-site energies/local magnetic fields, and their evolution with increasing system size in order to grant PSEM in experimentally-relevant spin-$1/2$ hardware. We show that the realisation of such local fields, beyond effective nearest-neighbour interactions, is vital for the fastest possible PSEM and furthermore their omission greatly reduces the compatibility with fault-tolerant quantum computing platforms. In summary, we present a trade-off of higher-engineering demands and markedly more robust and faster distribution of quantum information.
\end{abstract}
                              
\maketitle
\section{\label{intro}Introduction}
Since the formalisation of leveraging quantum systems for the purposes of computation was introduced \cite{10.1098/rspa.1985.0070}, significant effort has been devoted towards categorising  the essential characteristics \cite{DiVincenzo_2000} and eventual implementation \cite{Cirac_1995,Monroe_1995,Nakamura1999,Jaksch_2000,Vandersypen2001} of such a computer. Localised experimental and theoretical successes have led to global victories in increasing system sizes in which quantum phenomenon is observed, computation is executed, and interpretable information is extracted. These successes have led us to the so-called noisy intermediate scale quantum (NISQ \cite{Cheng_2023,Daka_2026}) device era \cite{Preskill_2018}, where additional degrees of freedom such as insulation difficulties, and/or loss of information to the extent it impedes practicality, have begun to surface \cite{Daka_2026,Xiang_2013,Cheng_2024}. 

Solid-state hardware in the form of spin systems with the exchange interaction may be exploited for universal quantum computation \cite{divincenzo_20002}. The transfer of low-loss information across quantum spin systems has been a pivotal quantum information processing task, in the pursuit of scaling quantum computing architectures, since the advent of utilising Heisenberg spin chains as a quantum-boosted channel for information transport/quantum state transfer (QST) \cite{Bose}. The capacity for such systems to aid and abet gate application, entanglement generation/distribution, and modular distribution of arbitrary pure states, is what has attracted attention. The last two decades have seen several protocols with varying degrees of success being proposed \cite{Christandl_2004,Albanese_2004,Karbach_2005,Apollaro_2012,Giorgi_2013,Lorenzo_2013,Christandl_2017,Bezaz_2025} and even implemented in a variety of real hardware \cite{Rao_2014,chapman2016,Li2018,Tian2020,Roy2025,Ge_2026}. Although there has been some limited work on high-fidelity state transfer (usually only between two outer nodes) in long-range (LR) interacting systems \cite{Kay_2006,Jex_2007,Gualdi_2008,Hermes_2020, Li_2025,Marshall_2026}, the vast majority of the analysis studied nearest-neighbour (NN) simulations, and treating emergent, and/or intentionally mediated, longer-range effects of various quantum platforms \cite{Cirac_1995,Johanning_2016,PhysRevApplied.17.034060,Sheremet2023,Defenu_2023} as an unavoidable transfer-fidelity taxation known to drastically impact the efficacy of the faithful transfer of information \cite{Ronke_2011}. 

In this work, we combine the approach originally proposed in \cite{Kay_2006,Jex_2007} with the analysis of total mirror inversion \cite{Albanese_2004,Karbach_2005} to yield perfect single-excitation mirroring (PSEM), which is any single-excitation state transferring to its respective mirrored position on the opposite half of the chain, across a variety of long-range interacting spin systems.
A distinct, though closely related, idea is that of perfect state transfer (PST), which in its effective meaning has been adapted and readapted in the broader literature. Broadly speaking, PST describes the transfer of a specific site encoding to another site along the chain with fidelity approaching (or exactly) unity. PST is a necessary but not sufficient condition for PSEM as, in general, PSEM requires not only one site \footnote{Often the outer nodes serve as the benchmark for transfer, as it utilises the full length of the system.} to transfer but the encoding of a single-excitation on any number of sites about one half of the chain to mirror. Furthermore, with PSEM all single-excitation encodings arrive at their respective mirrored side of the chain at the \textit{same} time. This is particularly useful for transferring of entangled-bits or `e-bits', as an example \cite{Lorenzo_2013}.  

We achieve PSEM through exploiting the linear spectrum, which sets the speed limit for perfect mirroring, and larger, long-range systems within the scale of current NISQ devices. We present the size $N-$dependent structural differences of systems capable of PSEM, over a range of interaction decay constant-$\alpha$ values of prospective technological interest within power-law decay and next-nearest-neighbour (NNN) models. In doing so, extending the investigation to larger, though not exceeding currently experimentally-accessible system sizes. The NNN solutions we present demonstrate a stark experimental advantage over general analytical solutions \cite{Christandl_2017}, in terms of total coherence time required for gate operation and experimental feasibility due to required energy differences. The power-law long-range results are particularly relevant for ion trap and Rydberg/cold-atom platforms, while the NNN results are promising for superconducting and waveguide architectures, where next-nearest-neighbour interactions can be unavoidable \cite{chapman2016,Xiang2023,Xiang2024,Lange_2026}. In such settings, experimentalists may be able to “turn a bug into a feature”, provided that these NNN couplings can also be tuned relative to the NN couplings \cite{PhysRevApplied.14.064016,Schulz_22}. We further present findings which demonstrate the necessity for on-site potentials for PSEM, even when the power-law decay of the interactions tend towards an effective nearest-neighbour model ($\alpha>5$). We then go on to discuss the numerical scaling and analytical asymptotic scaling of the quantum speed limit (QSL) in power-law LR models, using similar symmetric-spectral arguments originally used to derive the upper-bounds in NN settings \cite{Yung_2006,Kay_2022}. We do not find a difference in the QSL between NN and NNN models. Lastly, we examine the robustness of the power-law numerical LR solutions to positional and energy imprecision. We then propose a concrete experimental protocol, implementable with current hardware, that could improve the results of previous experiments, including those originally reported in Ref.~\cite{chapman2016}.  
\section{\label{model}Spin Model}
To describe the spin chain, we introduce the spin-$\frac{1}{2}$ so-called XX (also known as the isotropic XY) model
\begin{equation} \label{Hami}
    {\hat{H}_{\mathrm{XX}}}=
\frac{1}{2}\sum_{i < j}
J_{i,j}
\left(\sigma_i^x \sigma_j^x + \sigma_i^y \sigma_j^y \right)
+
\frac{1}{2}\sum_i B_i \sigma_i^z,
\end{equation}
where
\begin{equation}
J_{i,j}=\frac{J_0}{d_{i,j}^{\alpha}}
\label{Coupling}
\end{equation}
are the long-range coupling values dependent on the dimensionless separations of the site positions ($d_{i,j}= |d_i - d_j| = |i-j|$), and $B_i$ are the site($i$)-specific magnetic fields/on-site energies. Clearly, as we increase $\alpha$ within equation~\ref{Coupling}, it can be seen that equation~\ref{Hami} tends towards a NN chain, and as we decrease, we tend towards full network connectivity. Among the interaction exponents that naturally arise in quantum systems are $\alpha = 1$ (Coulomb) \cite{Welzel_2011,Johanning_2016}, $\alpha = 3$ (dipole--dipole) \cite{PhysRevLett.114.113002,PhysRevApplied.17.034060,Defenu_2023}, and $\alpha = 6$ (van der Waals) \cite{Defenu_2023,Adams_2019Rydberg,Zhao2023}. Once this latter value is exceeded, the system quickly assumes an effective nearest-neighbour model behaviour. It is also true, however, that experimentally the ratio of nearest-neighbour to next-nearest-neighbour couplings can be tuned \cite{Schulz_22,PhysRevApplied.14.064016}, allowing for regimes where the NNN interaction becomes a non-negligible fraction of the NN coupling while longer-range terms remain strongly suppressed. 
\section{Quantum State Transfer}\label{QST}
\subsection{Background}
For the QST procedure we consider, we adapt the benchmark which was originally proposed by Bose in Ref.~\cite{Bose}, which entails that an unknown single-qubit quantum state $\ket{\psi_s}=\cos\frac{\vartheta}{2}\ket{0}+\sin\frac{\vartheta}{2}e^{i \varphi}\ket{1}$ is encoded at any site of a spin chain and, exploiting the natural dynamics of the time-independent Hamiltonian, retrieved at time $t$ at some different location (the receiver site). The relative success of any spin chain configuration, of the kind described by equation~(\ref{Hami}), for the purposes of QST can be reasonably tracked by the so-called transfer fidelity, which is given by
\begin{align}
\label{eq:trans1}
    |f_i^j(t)|^2=\big\vert\bra{j}e^{-i \hat{H}_{XX} t}\ket{i}\big\vert^2=\big\vert\sum_{k=1}^N v_{ik}v_{jk}^*e^{-i \lambda_k t}\big\vert^2~,
\end{align}
where $\{\lambda_k, \ket{v_k}\}$, with $\ket{v_k}=\ket{v_{1k},v_{2k},\cdots,v_{Nk}}$, are the eigenvalues and eigenvectors of $\hat{H}_{\mathrm{XX}}$ when considering the single-excitation sector. We adopt the notation that $\ket{i}=
\ket{0_1}\otimes\cdots\otimes
\ket{1_i}
\otimes\cdots\otimes
\ket{0_N}$, with $  i \in \{1,\ldots,N\}$. Clearly, in order to witness a quantum state transferring `perfectly' we require equation~(\ref{eq:trans1}) to be equal to unity at some point in time $t_{PST,i}$. However, for PSEM to occur we require equation~(\ref{eq:trans1}) for all positions ($i$) and mirrored-positions ($j = N+1-i$) throughout the chain to be equal to unity at the \textit{same} `mirroring' time $t_{m}$. We therefore discuss PST, in perfect, site-specific transfer as being distinct, though a necessary condition for PSEM. From \cite{Christandl_2004}, we know (for $\alpha>>1$ and $J_0$ being constant) that the positions of the sites need to be spaced further at the edges and then gradually closer together towards the middle to reflect 
\begin{equation}
    J_{i,i+1} = J_0\sqrt{(N-i)i},
\end{equation}
for the resultant couplings.
This solution for PST in NN-systems yields a total state mirroring, due to its linear spectrum and mirror symmetry of the Hamiltonian.           
\label{T}

\subsection{Necessary and Sufficient Requirements}
With $| \psi_0 \rangle = | 1 \rangle$, for an $N$-site chain, PST between the end sites of the open-ended chain requires $| \psi_{final} \rangle = | N \rangle$, the `mirror image' of $|\psi_0\rangle$ with respect to the mirror operator, $M_{ij} = \delta_{i,N+1-j}$, or, in $N\times N$ matrix form
\begin{equation*}
    M = \begin{pmatrix}
  0 & 0 & \cdots & 0 & 1 \\
  0 & 0 & \cdots & 1 & 0 \\
  \vdots & \vdots & \ddots & \vdots & \vdots \\
  0 & 1 & \cdots & 0 & 0 \\
  1 & 0 & \cdots & 0 & 0 \\
\end{pmatrix}_{N \times N}
\label{M}.
\end{equation*}
For example,
\begin{equation}
\Hat{M}|1\rangle = \Hat{M}|1_10_20_3..0_N\rangle = \pm|0_10_20_3..1_N\rangle = |N\rangle.
\label{m}
\end{equation}
A first step towards ensuring PSEM dynamics is to choose a mirror-symmetric Hamiltonian. With $[\hat{H}_{XX},\Hat{M}]=0$, the energy eigenstates possess even or odd symmetry under $\Hat{M}$. Clearly, these symmetry conditions alone do not guarantee PST. In particular, homogeneous nearest-neighbour chains already exhibit deviations from PSEM for system sizes $N > 3$ \cite{Christandl_2004}, demonstrating that additional engineering is required. The second and final necessary step towards ensuring PSEM is to constrain the spectrum of the mirror-symmetric $H_\mathrm{XX}$. If the spectrum is such that at a later time $t_m$ (the mirror time) all the evolved odd symmetry eigenstates acquire an additional minus sign in comparison to all the even symmetry eigenstates, the effective evolution operator will be of the form
 $$ e^{-iH_\mathrm{XX}t_m/\hbar} = e^{-i \phi} \Hat{M},$$ for some global dynamical phase $\phi$. Thus, the initial state is mirrored and we may show from equation~(\ref{m}) that \begin{align}
f_i^{j}(t_m) &= |\langle N+1-i| e^{-i\hat{H}_{XX}t_m/\hbar}| i \rangle|^2 \notag \\
     &= |\langle N+1-i| e^{-i \phi} \Hat{M}| i \rangle|^2 \notag \\
     &= |\langle N+1-i|  N+1-i \rangle|^2 = 1.
     \label{Ftm}
\end{align}
Therefore, PSEM is achieved across any $N$-site chain with a mirror-symmetric Hamiltonian and corresponding eigenvalues that fulfil the required condition as outlined above. Another figure of merit for tracking the success of state transfer protocols, particularly relevant in the field of quantum communications, is the `average fidelity' \cite{Bose}. This metric is derived by averaging over all pure input states on the Bloch sphere and is given by
\begin{equation}
\langle F(t) \rangle = \frac{|f_1^N(t)|\cos(\nu)}{3} + \frac{|f_1^N(t)|^2}{6} + \frac{1}{2},
\label{AvF}
\end{equation}
where \( \nu = \text{arg}(f_1^N) \). To maximise \( \langle F(t)\rangle\), a system-specific global magnetic field \( \mathcal{B} \) can be applied across the chain, shifting each on-site energy to (\( B_i + \mathcal{B} \)), to ensure that \( \cos(\nu) = 1 \) in equation~(\ref{AvF}) \cite{Bose}.
A necessary condition for PSEM \cite{Kay_2010} is that the sequential (ordered) eigenvalue differences $\Delta\lambda_{k,k+1}$, are proportional to a pair-dependent odd integer ($Q_{k,k+1}$) such that
\begin{equation}
    \Delta\lambda_{k,k+1} = \frac{\pi}{t_m} Q_{k,k+1}.
    \label{eig}
\end{equation} for $k\in\{1,\ldots, N-1\}$.
The aforementioned requirement of mirror, centro-, or parity symmetry of the Hamiltonian matrix, together with equation~(\ref{eig}), guarantees total single-excitation state mirror inversion within XX spin chains.
\section{Inverse Eigenvalue Solver}
\label{IES}
Extending the approach introduced in \cite{Kay_2006,Jex_2007}, we develop a numerical, iterative procedure for PSEM by considering a one-dimensional quantum system of size $N$, described by a Hermitian, mirror-symmetric Hamiltonian $H(\beta)$, where the parameter vector
\begin{equation}
\beta =
\begin{cases}
(x_1,\ldots,x_{N/2},B_1,\ldots,B_{N/2}),
& N \text{ even},\\[4pt]
(x_1,\ldots,x_{(N-1)/2},
B_1,\ldots,B_{(N-1)/2},B_c),
& N \text{ odd}.
\end{cases}
\end{equation}
encodes spatial degrees of freedom and on-site fields. The spatial parameters define positive inter-site spacings via $d_{k,k+1} \equiv r_k = e^{x_k}$, which are used to construct a mirror-symmetric chain geometry. The physical spacings $\{r_i\}_{i=1}^N$ are generated recursively, enforcing mirror symmetry about the chain centre. The Hamiltonian is fully connected with algebraically decaying interactions,
\begin{equation}
H_{ij}(\beta) =
\begin{cases}
B_i, & i=j,\\[6pt]
\dfrac{J_0}{d_{i,j}^{\alpha}}, & i\neq j, 
\end{cases}
\end{equation}
where $\alpha$ is the power-law decay exponent. 
(We note that a completely connected graph with all equal couplings would be given by $\alpha=0$.) The on-site energies satisfy mirror symmetry $B_i = B_{N+1-i}$, with a possible unique central field $B_c = B_{(N+1)/2}$ for odd $N$. We target a linear spectrum such that all sequential eigenvalue pairs satisfy equation~(\ref{eig}), with \(Q_{k,k+1}=1\). The target eigenvalues are defined as
\begin{equation}
\tilde{\lambda}_k = \lambda_k^{(0)},
\end{equation}
where \(\lambda_k^{(0)}\) denotes the \(k\)-th eigenvalue of the desired linear spectrum. The eigenvalue mismatch is
\begin{equation}
\Delta \lambda_k = \tilde{\lambda}_k - \lambda_k(\beta),
\end{equation}
where \(\lambda_k(\beta)\) are the eigenvalues of \(H(\beta)\). First-order perturbation theory \cite{cohen_1992} defines the sensitivity matrix
\begin{equation}
J_{k\mu} =
u_k^\dagger
\left(
\frac{\partial H}{\partial \beta_\mu}
\right)
u_k
\end{equation}
where $\mu$ tracks individual perturbations.
For spatial parameters, derivatives take the form
\begin{equation*}
\frac{\partial H_{ij}}{\partial x_k}
=
-\alpha (d_{i,j})^{-\alpha-1}
\frac{\partial (d_{i,j})}{\partial r_k}\, r_k,
\end{equation*}
with geometric sensitivity 
\begin{equation*}
\frac{\partial d_{i,j}}{\partial r_k}
=
\begin{cases}
1, & k \text{ lies between } i \text{ and } j,\\
0, & \text{otherwise}.
\end{cases}
\end{equation*}
and for on-site fields $\frac{\partial H}{\partial B_i} = |i\rangle \langle i| +|N+1-i\rangle\langle N+1-i|$.
The inverse eigenvalue problem is therefore solved iteratively  using a damped Gauss-Newton update \cite{Nocedal_2006}
\begin{equation}
\Delta \beta =
\left(J^\top J + \eta(N) I\right)^{-1}
J^\top \Delta \lambda,
\end{equation}
with regularisation and damping chosen as
\begin{equation*}
\eta(N) = 10^{-6}\frac{N}{\xi},
\qquad
\gamma(N) = 0.05\frac{\xi}{N}.
\end{equation*}
where $\xi$ was set as a damping constant $\sim 10$.
The parameter vector is updated iteratively according to
\[
\beta_{\mathrm{new}}
=\beta_{\mathrm{old}}+\gamma(N)\Delta\beta,
\]
where \(\beta_{\mathrm{old}}\) and \(\beta_{\mathrm{new}}\) denote the parameter vectors before and after the update, respectively, and the iteration was set to terminate when $\|\Delta \lambda\|_2 < 10^{-8}$ or after a fixed number of iterations ($\sim10^3$) in the event of no convergence. The execution of this iterative procedure was conducted over a range $N\sim4-100$ across $\alpha \in [1,3,6]$ for LR models described by XX Hamiltonian matrices. As an example, here and in the following, we explicitly provide results for the $N=8$ power-law LR Hamiltonian
\begin{equation} \setlength{\arraycolsep}{1pt} \resizebox{0.927\textwidth}{!}{$ H_1= \begin{pmatrix} B_1 & \frac{1}{r_1^\alpha} & \frac{1}{(r_1+r_2)^\alpha} & \frac{1}{(r_1+r_2+r_3)^\alpha} & \frac{1}{(r_1+r_2+r_3+r_4)^\alpha} & \frac{1}{(r_1+r_2+2r_3+r_4)^\alpha} & \frac{1}{(r_1+2r_2+2r_3+r_4)^\alpha} & \frac{1}{(2r_1+2r_2+2r_3+r_4)^\alpha} \\[6pt] \frac{1}{r_1^\alpha} & B_2 & \frac{1}{r_2^\alpha} & \frac{1}{(r_2+r_3)^\alpha} & \frac{1}{(r_2+r_3+r_4)^\alpha} & \frac{1}{(r_2+2r_3+r_4)^\alpha} & \frac{1}{(2r_2+2r_3+r_4)^\alpha} & \frac{1}{(r_1+2r_2+2r_3+r_4)^\alpha} \\[6pt] \frac{1}{(r_1+r_2)^\alpha} & \frac{1}{r_2^\alpha} & B_3 & \frac{1}{r_3^\alpha} & \frac{1}{(r_3+r_4)^\alpha} & \frac{1}{(2r_3+r_4)^\alpha} & \frac{1}{(r_2+2r_3+r_4)^\alpha} & \frac{1}{(r_1+r_2+2r_3+r_4)^\alpha} \\[6pt] \frac{1}{(r_1+r_2+r_3)^\alpha} & \frac{1}{(r_2+r_3)^\alpha} & \frac{1}{r_3^\alpha} & B_4 & \frac{1}{r_4^\alpha} & \frac{1}{(r_3+r_4)^\alpha} & \frac{1}{(r_2+r_3+r_4)^\alpha} & \frac{1}{(r_1+r_2+r_3+r_4)^\alpha} \\[6pt] \frac{1}{(r_1+r_2+r_3+r_4)^\alpha} & \frac{1}{(r_2+r_3+r_4)^\alpha} & \frac{1}{(r_3+r_4)^\alpha} & \frac{1}{r_4^\alpha} & B_4 & \frac{1}{r_3^\alpha} & \frac{1}{(r_2+r_3)^\alpha} & \frac{1}{(r_1+r_2+r_3)^\alpha} \\[6pt] \frac{1}{(r_1+r_2+2r_3+r_4)^\alpha} & \frac{1}{(r_2+2r_3+r_4)^\alpha} & \frac{1}{(2r_3+r_4)^\alpha} & \frac{1}{(r_3+r_4)^\alpha} & \frac{1}{r_3^\alpha} & B_3 & \frac{1}{r_2^\alpha} & \frac{1}{(r_1+r_2)^\alpha} \\[6pt] \frac{1}{(r_1+2r_2+2r_3+r_4)^\alpha} & \frac{1}{(2r_2+2r_3+r_4)^\alpha} & \frac{1}{(r_2+2r_3+r_4)^\alpha} & \frac{1}{(r_2+r_3+r_4)^\alpha} & \frac{1}{(r_2+r_3)^\alpha} & \frac{1}{r_2^\alpha} & B_2 & \frac{1}{r_1^\alpha} \\[6pt] \frac{1}{(2r_1+2r_2+2r_3+r_4)^\alpha} & \frac{1}{(r_1+2r_2+2r_3+r_4)^\alpha} & \frac{1}{(r_1+r_2+2r_3+r_4)^\alpha} & \frac{1}{(r_1+r_2+r_3+r_4)^\alpha} & \frac{1}{(r_1+r_2+r_3)^\alpha} & \frac{1}{(r_1+r_2)^\alpha} & \frac{1}{r_1^\alpha} & B_1 \end{pmatrix} $}. \label{LRN=8} \end{equation}
 One may notice that equation~(\ref{LRN=8}) presents a mirror-symmetric/persymmetric matrix and therefore if such a matrix is found to have a linear (or simply commensurate, equation~(\ref{eig})) spectrum, PSEM is guaranteed. 
\section{Results}
In this section, we present over a range of $ \alpha $, PSEM localised from any $i^{th}$ site in the chain to its mirrored $(N-i+1)^{th}$ site for power-law LR and NNN chains, for system sizes of immediate and near-term experimental interest ($N \sim10 -10 ^ 2$). All protocols were independently cross-validated numerically, with convergence verified to a tolerance of \(10^{-8}\).    
\subsection{Power-Law Long-Range (LR) Models}
\begin{figure}[h!]
    \centering
    \includegraphics[width=0.6\linewidth]{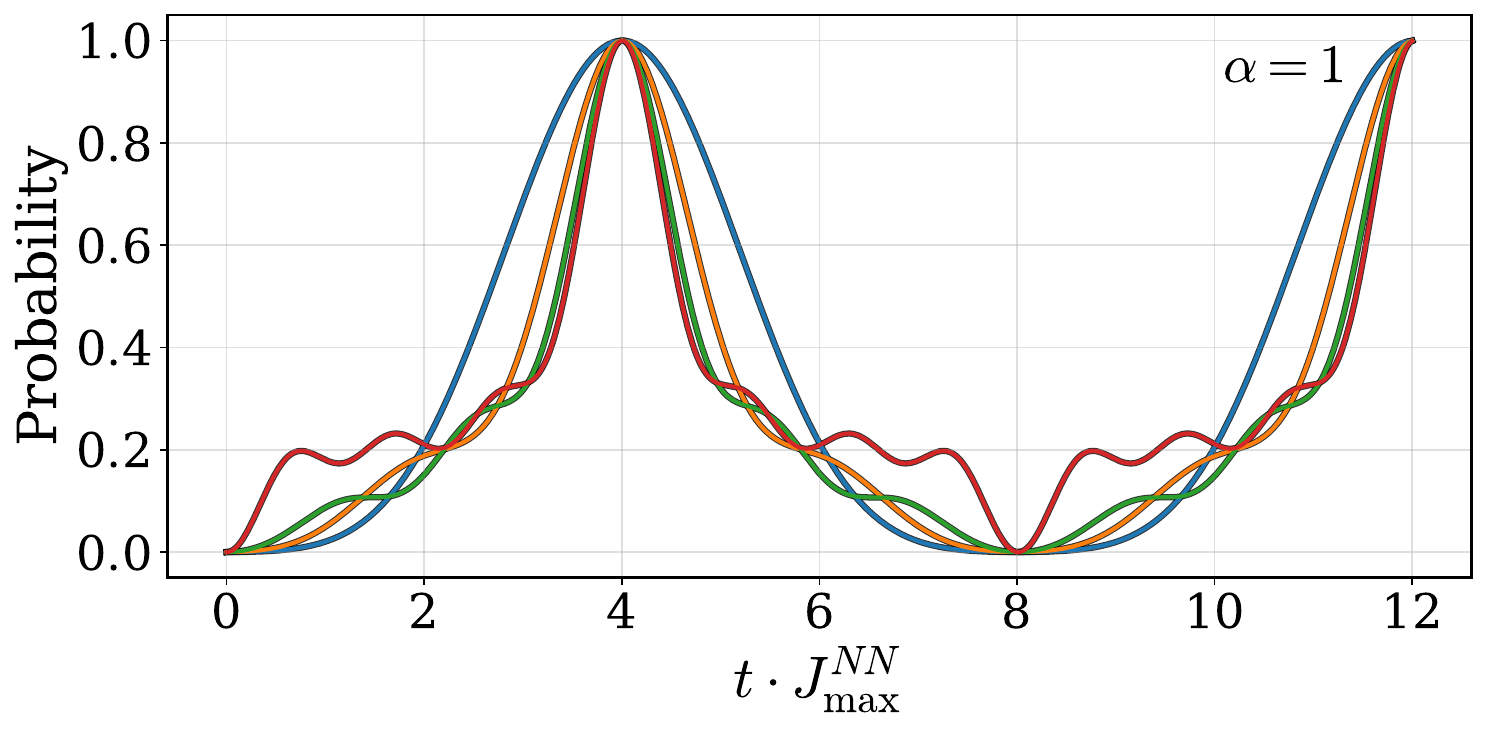}\\[-11mm]
\includegraphics[width=0.6\linewidth]{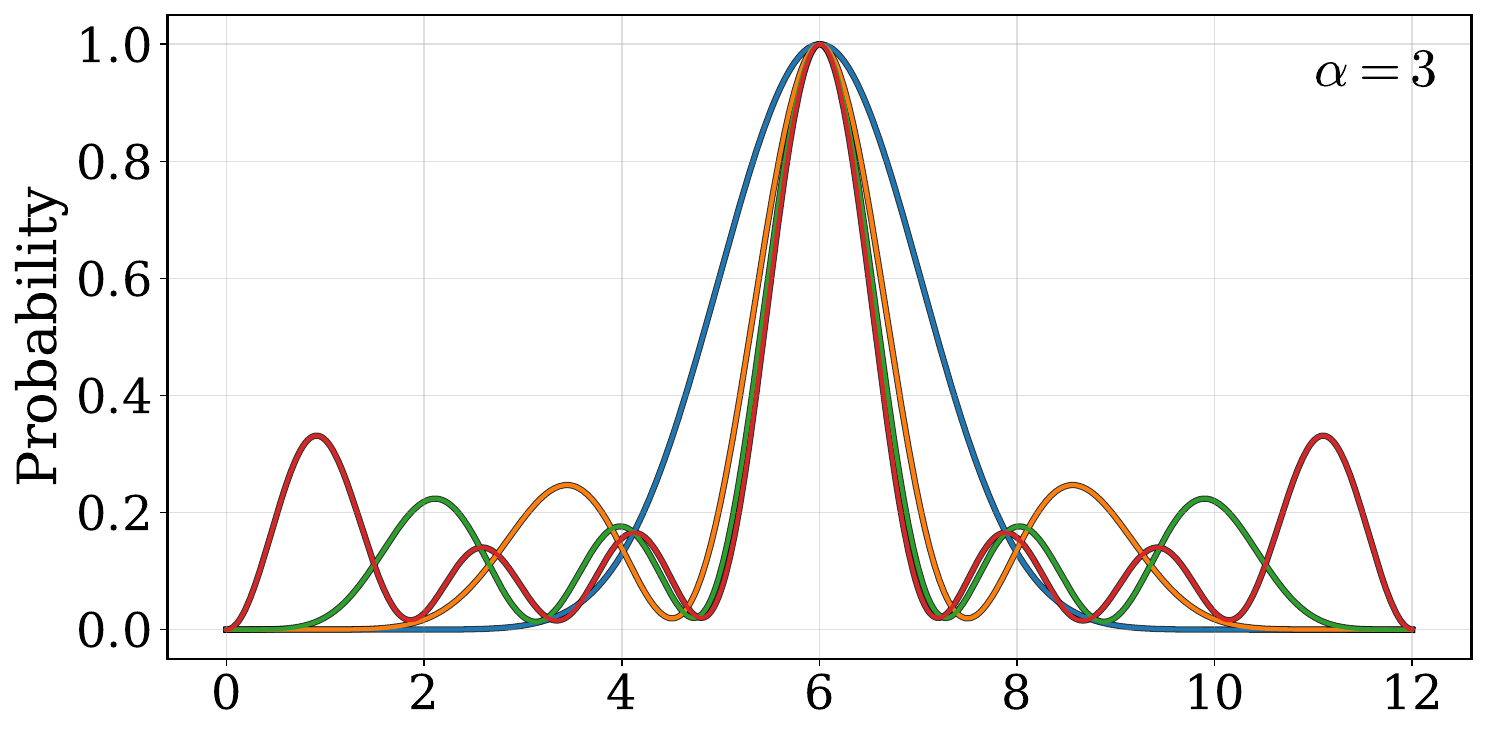}\\[-5mm]

\includegraphics[width=0.6\linewidth]{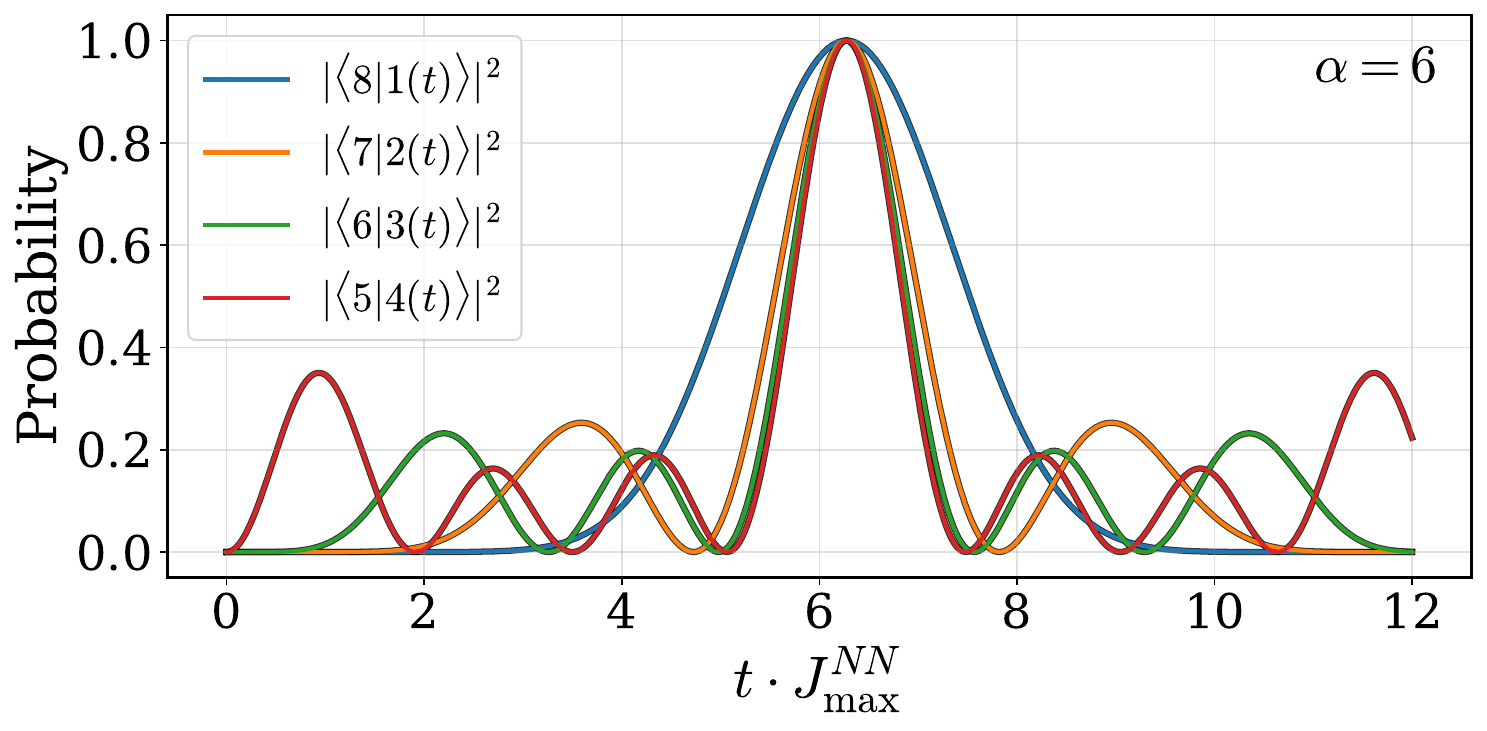}
    \caption{Transfer fidelity dynamics, normalised by the maximal nearest-neighbour ($NN$) coupling ($J^{NN}_{\mathrm{max}}$), for each site $i$ and mirrored position $j$ within the N = 8 chain described by equation~(\ref{LRN=8}), for naturally emerging $\alpha$-values. Numerical values corresponding to the above fidelity dynamics for $r_{i}(\alpha)$ and $B_i(\alpha)$ may be found in Table.~\ref{N=8}.}
    \label{fig_1}
\end{figure}

\begin{figure}[H]
    \centering
    \includegraphics[width=1.0\linewidth]{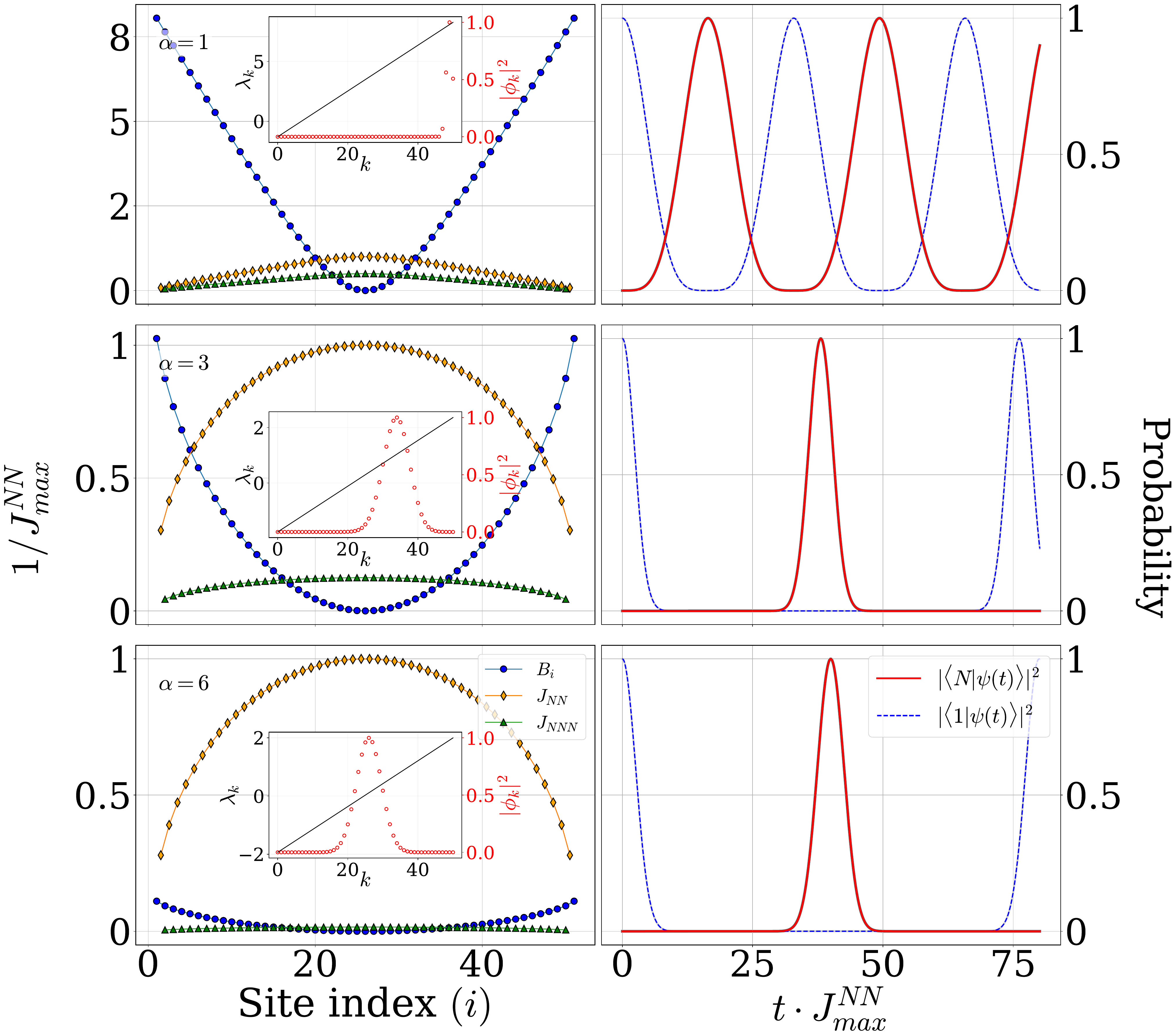}
    
    \caption{PSEM-$N=51-$site chain for varying $\alpha$=1 (top), 3 (central), and 6 (bottom), resultant from a completely linear spectrum (inset figures), and mirror symmetric Hamiltonian of the kind described by equation~(\ref{LRN=8}). The inset of the left column shows the linearity of the spectrum (black line, left axis) for the fifty eigenmodes, whilst the right-axis shows the initial-state localisation through the spectral weights $\phi_k= v_{1k}$ (equation~\ref{eq:trans1}) in $k$-space (red dots, right axis) to demonstrate the distribution of dynamically-effective modes when transferring $|1\rangle\rightarrow|N\rangle.$}
    \label{fig_2}
\end{figure}

\begin{figure}[H]
    \centering
    \includegraphics[width=0.85\linewidth]{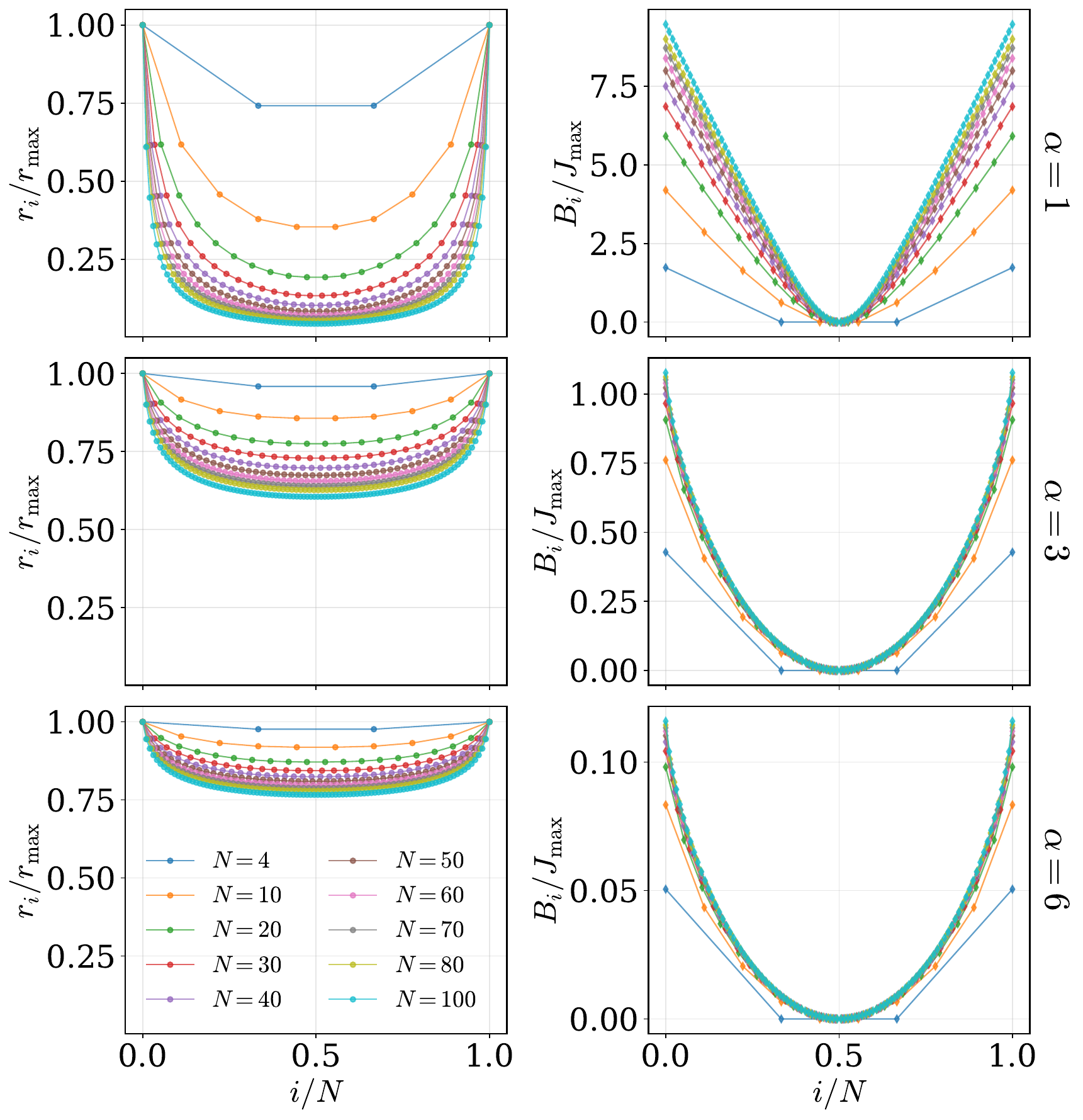}
    \caption{Normalised positions (left column) and ratio between local magnetic fields/on-site energies (right column) trends along the normalised site positions ($x$-axis) throughout the chain for varying alpha-valued LR PSEM solutions. All results are for $N=4-100$.}
    \label{fig_3}
\end{figure}

Table~\ref{N=8} shows the mirror-symmetric positions and on-site energies required for an $N$=8-site power-law LR chain with different decay-constant values $\alpha$ of the kind shown in equation~(\ref{N=8}), resultant from the inverse eigenvalue procedure outlined in the previous section. The corresponding transfer dynamics and corresponding exhibition of PSEM are presented in figure~\ref{fig_1}. The PSEM is enabled by the relative modulation of the positions and on-site energies of the model corresponding to the chosen $\alpha$ value. We see clearly in figure~\ref{fig_1} that not only is there PSEM in these solutions, but that there is a non-trivial difference in the  mirroring time $t_m$, where the lower-$\alpha$ solutions are markedly faster. From figure~\ref{fig_2}, we can see this speed advantage increase and saturate with increasing $N$ and decreasing $\alpha$, tending towards the NN chain limit (which requires no on-site energy variation). We note that the Coulomb interaction ($\alpha=1$) attains two full mirroring cycles during the time that the dipole-dipole and van Der Waals interaction present just one. We also observe notable differences in the transfer dynamics due to the connectivity of the network amongst the varying $\alpha$-examples. Namely for the $\alpha = 1$ there is consistent findings of few-mode coherent oscillations \cite{Marshall_2026}, as only 3-4 modes appear to participate in the dynamics of transferring the excitation from the first to the last site. Additionally, there is no ``early-state exclusion" \cite{Bailey2025} (i.e. we get $\langle 1|1(t)\rangle\neq0$ when $\langle N|1(t)\rangle>0$), which is often present in the case where there is wavepacket propagation, due to the excitation maintaining (mostly) a pronounced peak along the system \cite{Apollaro_2012}, and leaving the first site completely before recombining on the last. Due to the greater topological complexity of these long-range models, the linear and symmetric spectrum $\lambda_k=-\lambda_{N+1-n}$ no longer guarantees zero on-site energy contributions \cite{Vinet_2012}. We observe that the degree of influence on the dynamics of the on-site energies vary strongly across $\alpha-$values, with $\alpha=1$ exhibiting the strongest on-site energy contributions. Whilst for $\alpha=6$ we have more modest corrections in the on-site energies and the energy landscape is coupling-dominated.
\begin{table}[t]
\small
\caption{Half-chain mirror-symmetric spacings $r_{i}$ and shifted on-site energies $B_i \rightarrow B_i - \min(B_i)$ values (rounded) for PSEM within $N=8$-site power-law LR chains (equation~(\ref{LRN=8})). Only $i=1\ldots4$ are shown; full chains follow $r_{i} = r_{N-1-i}$ and $B_i = B_{N-1-i}$.}
\centering
\begin{ruledtabular}

\setlength{\tabcolsep}{40pt}
\begin{tabular}{c|c|c|c|c}
$H_{ij}$ & $\alpha=1$ & $\alpha=3$ & $\alpha=6$ & $\alpha=100$ \\
\hline
$r_{1}$ & 6.422 & 1.386 & 1.176 & 1.0098 \\

$r_{2}$ & 3.979 & 1.277 & 1.124 & 1.0070 \\

$r_{3}$ & 3.028 & 1.235 & 1.104 & 1.0059 \\

$r_{4}$ & 2.745 & 1.223 & 1.098 & 1.0056 \\
\hline
$B_1$ & 1.315 & 0.381 & 0.043 & 0 \\

$B_2$ & 0.765 & 0.169 & 0.019 & 0 \\

$B_3$ & 0.295 & 0.054 & 0.006 & 0 \\

$B_4$ & 0.000 & 0.000 & 0.000 & 0 \\

\end{tabular}
\end{ruledtabular}
\label{N=8}
\end{table}
As shown in figure~\ref{fig_3} for $N = 4-100$, the range of on-site energy strengths across the power-law LR models scales inversely with the decay exponent $\alpha.$ Similarly, the trend in the normalised, maximal spacing differences $r_i/r_{max}$ exhibit a wider range of values towards the centre of the chain as well in the case of lower$-\alpha$ chains. Both of these smooth, parabolic adjustments appear to be mandatory in adjusting the eigenvalues to maintain spectral-linearity, consistent with the NN picture (with no on-site energies required).  
\subsection{Next-nearest-neighbour (NNN) Models}
\noindent Another form of long-range models of specific technological interest are NNN models. They take the form of a mirror-symmetric pentadiagonal matrix, such as, for $N=8$
\begin{equation}
H_2 =
\begin{pmatrix}
B_1 & \frac{1}{r_1^\alpha} & \frac{1}{(r_1+r_2)^\alpha} & 0 & 0 & 0 & 0 & 0 \\
\frac{1}{r_1^\alpha} & B_2 & \frac{1}{r_2^\alpha} & \frac{1}{(r_2+r_3)^\alpha} & 0 & 0 & 0 & 0 \\
\frac{1}{(r_1+r_2)^\alpha} & \frac{1}{r_2^\alpha} & B_3 & \frac{1}{r_3^\alpha} & \frac{1}{(r_3+r_4)^\alpha} & 0 & 0 & 0 \\
0 & \frac{1}{(r_2+r_3)^\alpha} & \frac{1}{r_3^\alpha} & B_4 & \frac{1}{r_4^\alpha} & \frac{1}{(r_3+r_4)^\alpha} & 0 & 0 \\
0 & 0 & \frac{1}{(r_3+r_4)^\alpha} & \frac{1}{r_4^\alpha} & B_4 & \frac{1}{r_3^\alpha} & \frac{1}{(r_2+r_3)^\alpha} & 0 \\
0 & 0 & 0 & \frac{1}{(r_3+r_4)^\alpha} & \frac{1}{r_3^\alpha} & B_3 & \frac{1}{r_2^\alpha} & \frac{1}{(r_1+r_2)^\alpha} \\
0 & 0 & 0 & 0 & \frac{1}{(r_2+r_3)^\alpha} & \frac{1}{r_2^\alpha} & B_2 & \frac{1}{r_1^\alpha} \\
0 & 0 & 0 & 0 & 0 & \frac{1}{(r_1+r_2)^\alpha} & \frac{1}{r_1^\alpha} & B_1
\end{pmatrix}
\label{13}
\end{equation}
where alpha describes the `decay' from nearest-neighbour to next-nearest-neighbour strengths. If such a model were to be realised in actual power-law platforms, we would expect the NNN-coupling to decay quite rapidly ($\alpha\geq5$), so that couplings beyond NNN would not have a substantial influence on the dynamics. However, we will present varying decay-values of $\alpha=2.3-6.3$ within this model as well, to cover a range of experimentally-realised NN-to-NNN couplings $J_{max}^{\mathrm{NNN}}/J_{max}^{\mathrm{NN}}\simeq[0.2-0.01]$ \cite{chapman2016,Schulz_22, Xiang2023}. Table~\ref{N=8NNN} presents the spacings and on-site profiles required for $N=8$ NNN chain across decreasing NN-to-NNN coupling ratios. We note that, unlike the genuine power-law long-range $\alpha-$models, there are no significant differences in the mirroring times between the solutions shown. 

\begin{figure}[H]
    \centering
    \includegraphics[width=0.85\linewidth]{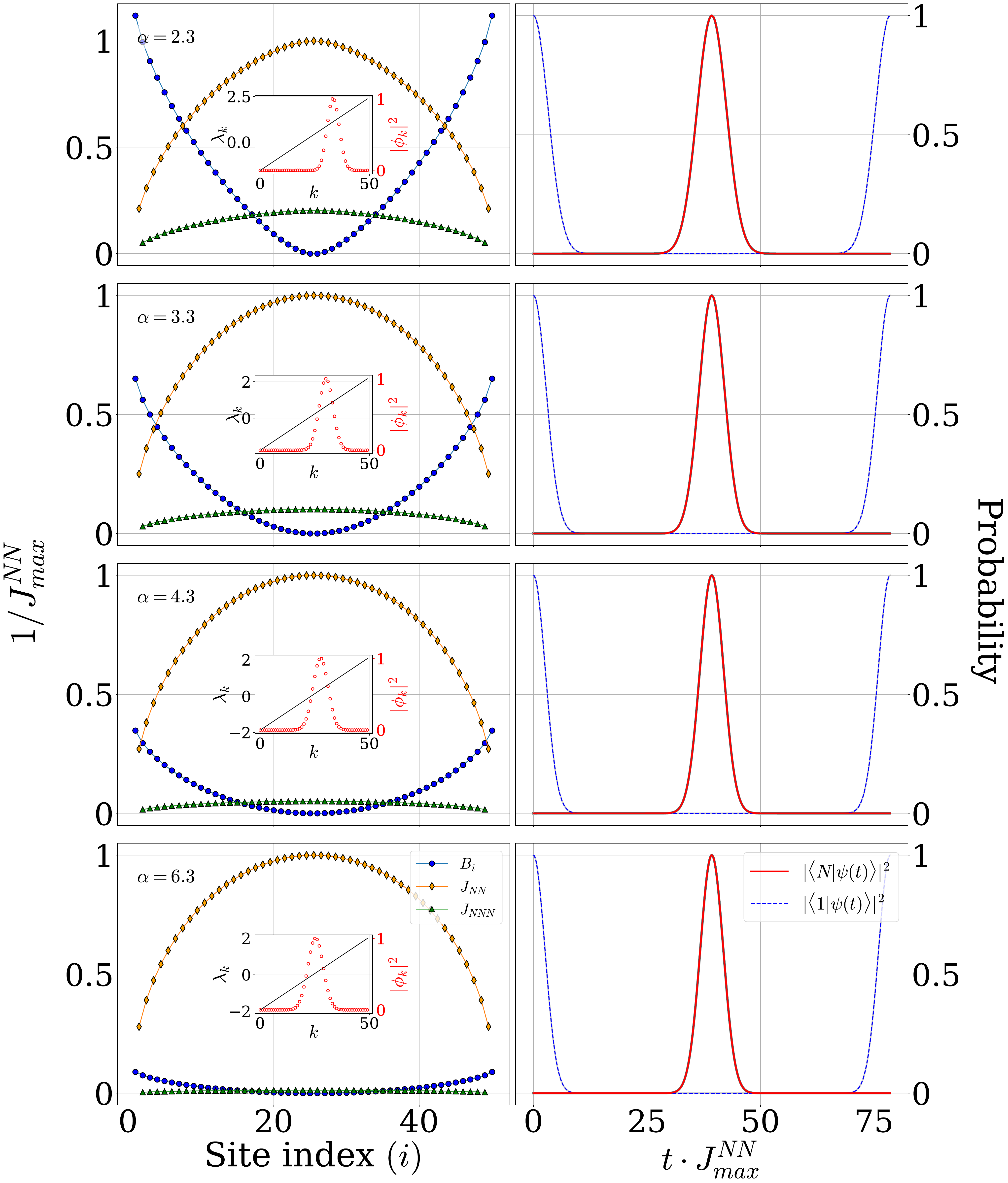}
    \caption{Coupling and on-site energy structure (left column) corresponding to specifically chosen decay constant $\alpha$-values for an $N=$50-site capable of PSEM due to the linearity of their spectrum, with the corresponding transfer dynamics (right column). The inset of the left column shows the linearity of the spectrum (black line, left axis) for the fifty eigenmodes, whilst the right-axis shows the initial-state localisation $\phi_k= v_{1k}$ (equation~\ref{eq:trans1}) in $k$-space (red dots, right axis) to demonstrate the distribution of dynamically-effective modes when transferring $|1\rangle\rightarrow|N\rangle.$ All solutions occur at the same time (in units of $1/J^{NN}_{\mathrm{max}})$ of $50\pi/4$ which is the speed limit for nearest-neighbour systems.}
    \label{fig_4}
\end{figure}

\begin{figure}[H]
    \centering
    \includegraphics[width=0.9\linewidth]{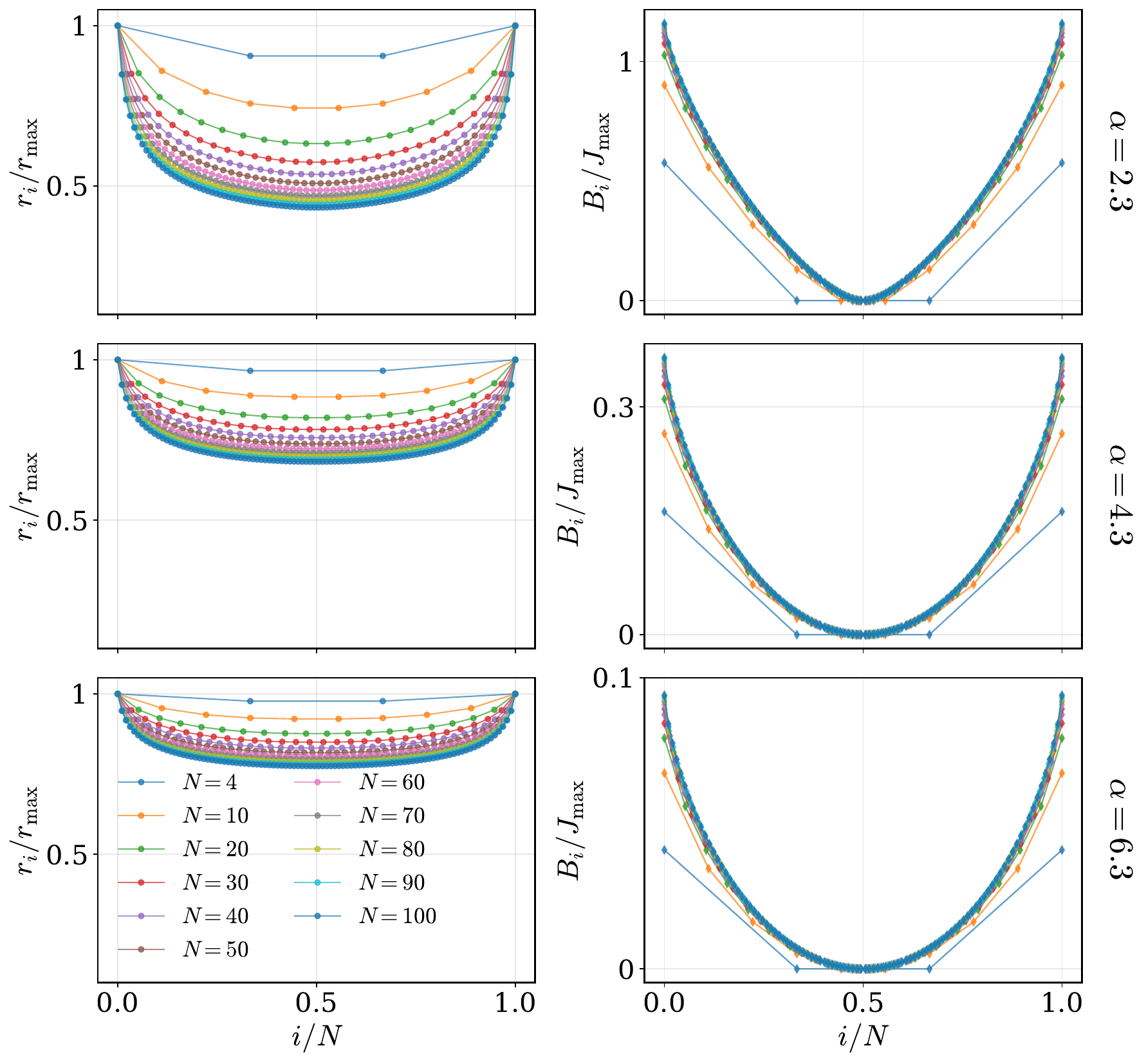}
    \caption{How the normalised positions (left column) and local magnetic fields/on-site energies (right column) change along the normalized site positions throughout the truncated LR-NNN-chains (of the kind of equation~(\ref{N=8NNN}) for varying alpha-values. All results are for $N=4-100$ and present $\alpha$-values which correspond to a range of experimentally-realised NN-to-NNN couplings $J_{max}^{\mathrm{NNN}}/J_{max}^{\mathrm{NN}}\simeq[0.2-0.01].$ }
    \label{fig_7}
\end{figure}

\begin{figure}[H]
    \centering
    \includegraphics[width= 0.4\linewidth]{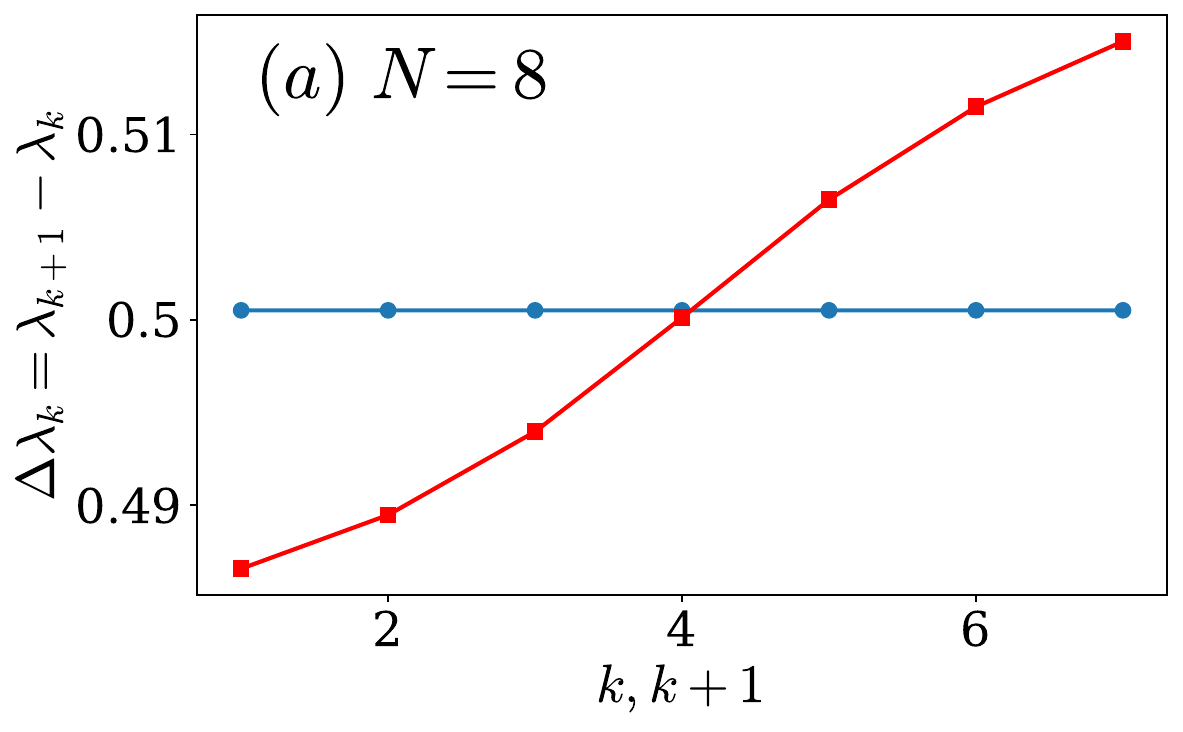}
    \includegraphics[width=0.4\linewidth]{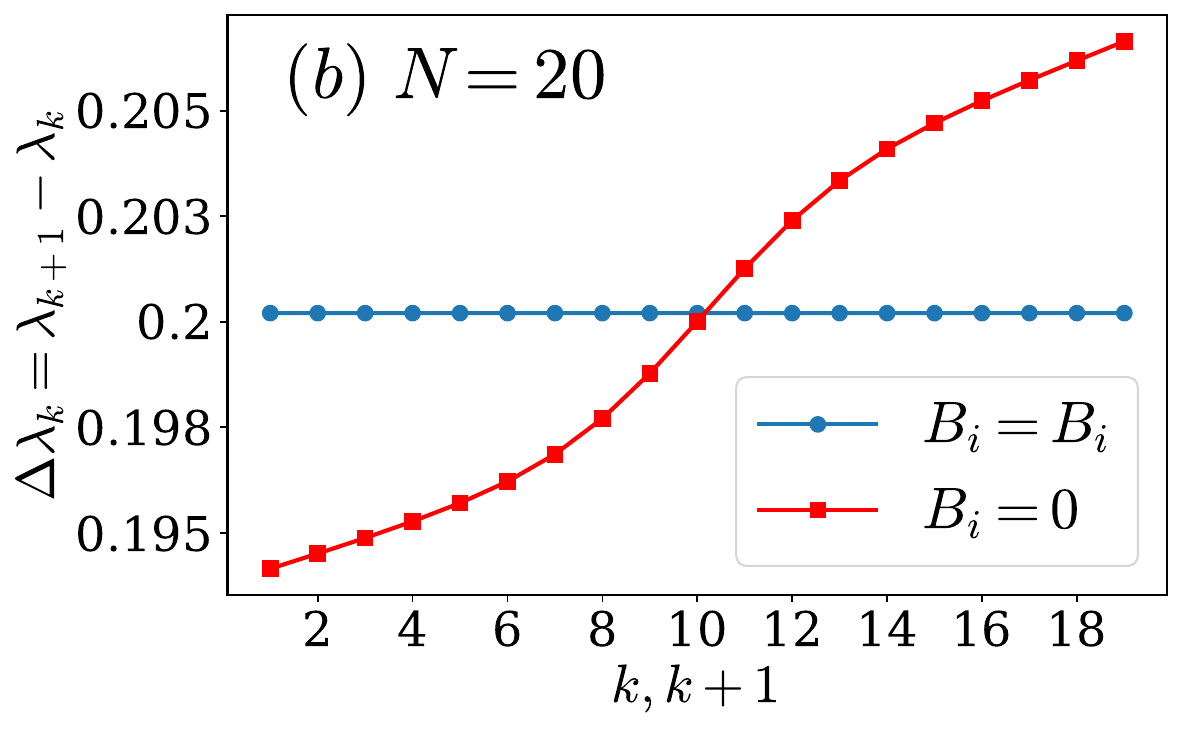}
    \includegraphics[width=0.8\linewidth]{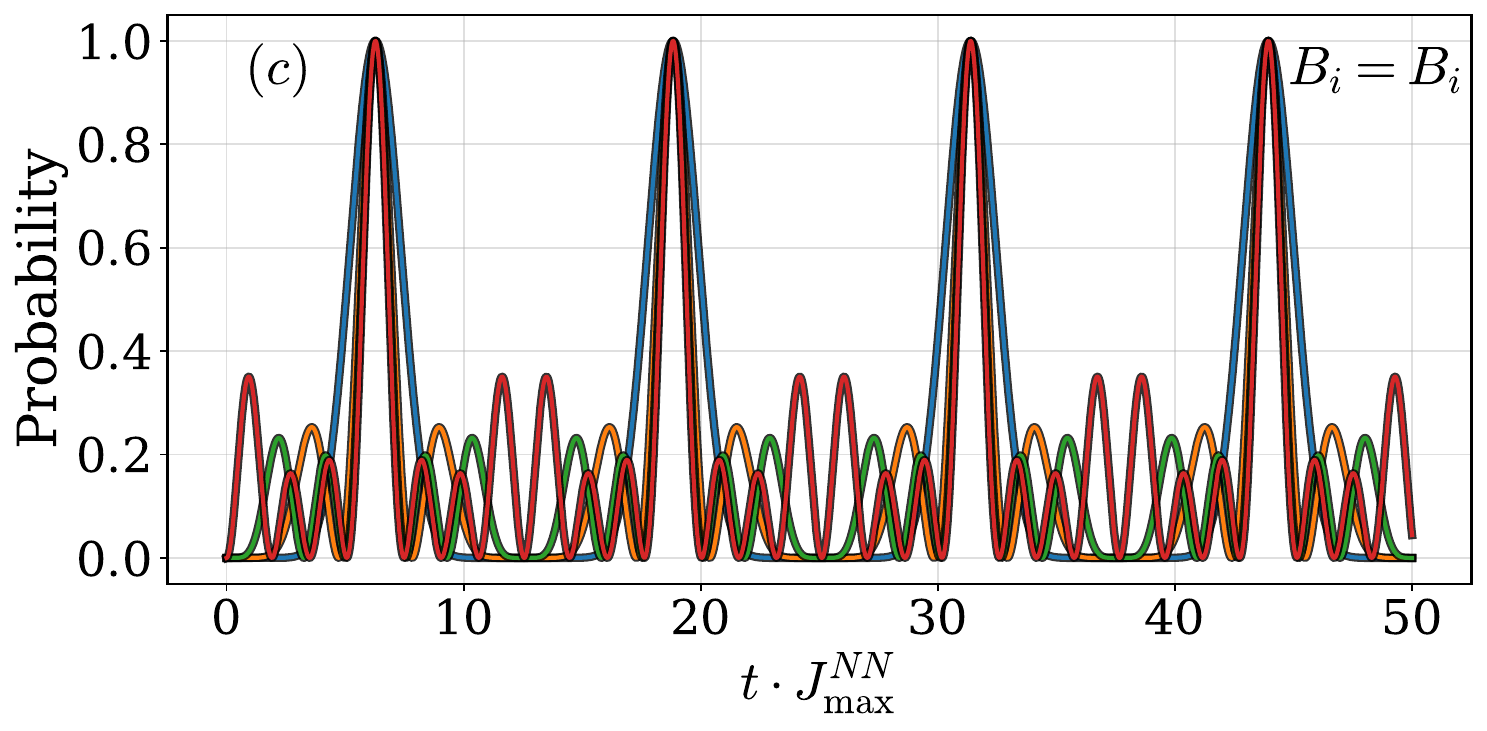}\\[-14mm]
\includegraphics[width=0.8\linewidth]{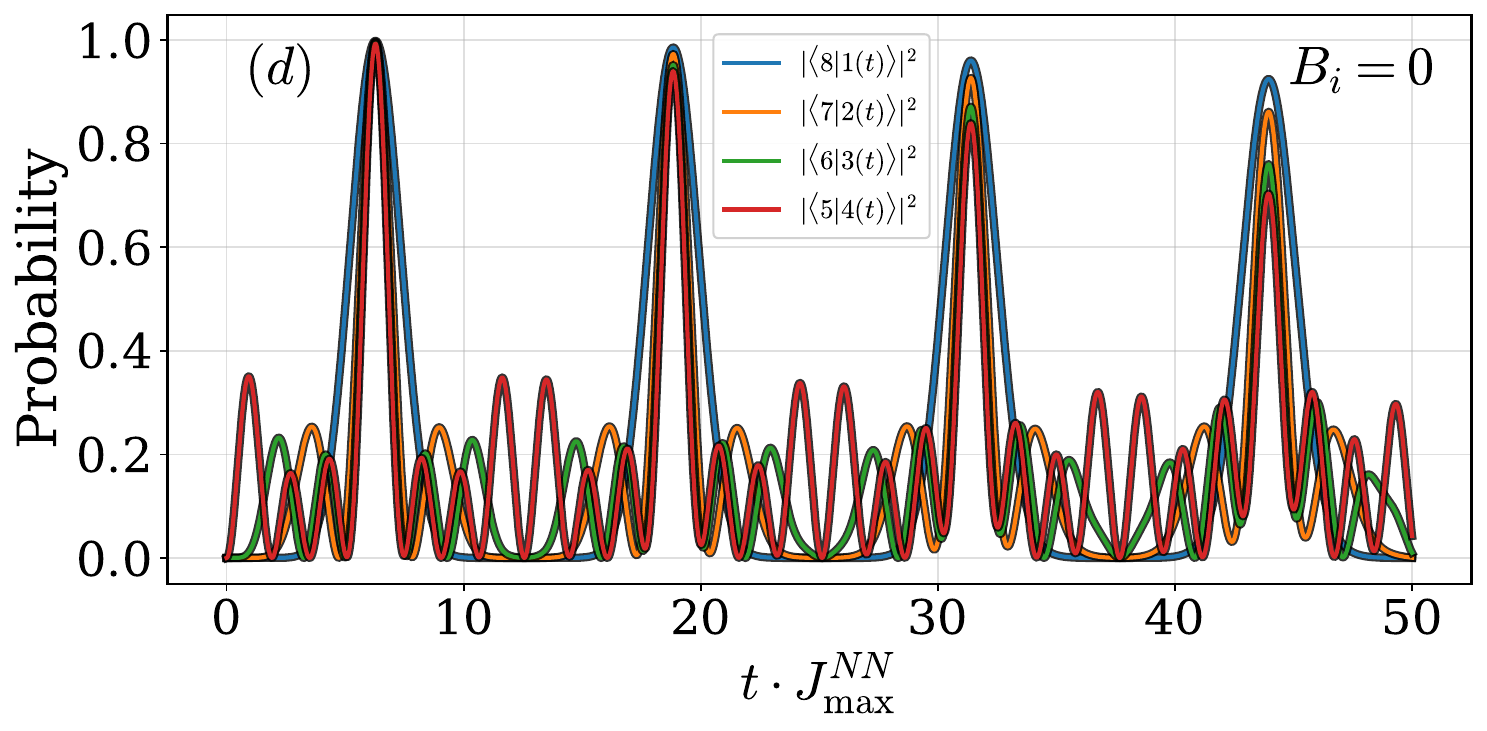}
    \caption{Eigenvalue spacing differences ($\Delta\lambda_k$) in (a) and (b), and transfer fidelity dynamics (c) and (d) for each site $i$ and mirrored position $j=N+1-i$ for an $N=8$-XX chain ($N=20$ only for (b)) described by equation~(\ref{13}) for $\alpha=6.3$ (positions and on-site energy profile in Table.~\ref{N=8NNN}) NNN-to-NN coupling ratio $\sim10^{-1}\cdot J^{NN}_{max}$. When on-site energies are switched off entirely it not only impacts the maximal fidelity of first peak (mean maximal fidelity value between all sites $i$ and mirrored site is 0.99577 instead of 1.0) it causes a decay in efficiency of each subsequent attempt, which is consistent with QPST \cite{Bezaz_2025,Nelmes_2026} rather than PSEM.}
    \label{fig_5}
\end{figure}

Indeed, figure~\ref{fig_4} highlights that despite significant structural (Hamiltonian) differences between the solutions, their mirroring times due to their linear spectra are nearly identical to each other and closely align with the speed limit in the NN context. 
Figure~\ref{fig_7} presents structural similarities in the trends of the required spacings and on-site energies as $N$ increases for different coupling ratios in comparison to the power-law LR solutions (figure~\ref{fig_3}). Namely, the spacings in lower-alpha settings must have larger differences coming from the edges towards the centre. As well as the on-site energies requiring a larger maximum-to-minimum variation in lower-$\alpha$ solutions. Figure~\ref{fig_5} (a) and (b) presents the differences in the spectrum between the PSEM solution and when the on-site energies are completely omitted for $\alpha=6.3$, within system sizes $N=8$ and 20, respectively. Figure~\ref{fig_5} (c) and (d) demonstrates that even for $\alpha = 6.3$, for which  $J_{max}^{NNN}/J_{max}^{NN}\sim0.01$, a non-negligible (in terms of the maximal coupling) harmonic on-site profile (see table~\ref{N=8NNN}) is still required in order to enable PSEM. In its absence, the state transfer efficacy diminishes  with worsening over longer time frames. These on-site energy corrections within the LR solutions are antithetical to homogeneously coupled nearest-neighbour systems which effect quasi-perfect state transfer (QPST) due to harmonic on-site profiles alone \cite{Bezaz_2025, Nelmes_2026}. In such systems, in order to transition from QPST to PSEM, a similarly small increase of the couplings (instead of the on-site energies slightly increasing at the edges within the LR systems) towards the centre is required \cite{Nelmes_2026}. We find worsening in the transfer fidelity and the systems ability to effect PSEM for larger system sizes if on-site energies corresponding to the ideal PSEM solutions are omitted (see figure~\ref{fig_10} and table~\ref{tab3} in Appendix \ref{reduction}). However, in the case of $B_i=0$, the reduced transfer-fidelity peaks for the site-to-mirror-site pairs occur, to a very good approximation, at the same times as in the optimal case, at least over the first revivals (figure~\ref{fig_5}(c) and (d)). As shown in figure~\ref{fig_5}(a) and (b), this can be attributed to the shifts of the eigenvalues being relatively small and, to a good approximation, symmetric with respect to the centre of the spectrum. The corresponding eigenvalue spacings differ by only a few per cent from the original (less than 3\%) and the Hamiltonian's mirror symmetry is preserved. Although the small spectrum asymmetry theoretically leads to accumulated dephasing at longer times, the timing of the first revivals remains robust, even when the on-site energies cannot be controlled experimentally.
\begin{table}
\centering
\caption{Half-chain mirror-symmetric spacings $r_{i}$ and shifted on-site energies $B_i \rightarrow B_i - \min(B_i)$ values (rounded) for PSEM within $N=8$-site NNN chain (equation~(\ref{13})). Only $i=1\ldots4$ are shown; full chains follow $r_{i} = r_{N-1-i}$ and $B_i = B_{N-1-i}$.}
\begin{ruledtabular}

\setlength{\tabcolsep}{40pt}
\begin{tabular}{c|c|c|c|c}

$H_{ij}$ & $\alpha=2.3$ & $\alpha=3.3$ & $\alpha=4.3$ & $\alpha=6.3$ \\
\hline

$r_1$ & 1.631 & 1.364 & 1.258 & 1.167 \\
$r_2$ & 1.409 & 1.246 & 1.179 & 1.118 \\
$r_3$ & 1.311 & 1.199 & 1.148 & 1.099 \\
$r_4$ & 1.275 & 1.185 & 1.139 & 1.093 \\

\hline

$B_1$ & 0.482 & 0.270 & 0.140 & 0.035 \\
$B_2$ & 0.257 & 0.126 & 0.061 & 0.015 \\
$B_3$ & 0.105 & 0.043 & 0.019 & 0.005 \\
$B_4$ & 0.000 & 0.000 & 0.000 & 0.000 \\

\end{tabular}
\label{N=8NNN}
\end{ruledtabular}
\end{table}
A general (in $N$) analytical solution of PSEM in NNN models has been derived \cite{Christandl_2017}. However, the resultant energy profile of the couplings and on-site energies create very large energy differences between the terminal sites and the central ones, creating a fundamental problem for experimental realisations. In addition to this, the time taken when normalising the dynamics with respect to the maximal nearest-neighbour couplings (the smallest maxima of the couplings and/or on-site energies in this system) requires significantly longer time frames than in the case of exclusive nearest-neighbour interactions \cite{Faria_2025}. This is due to non-linearity of the analytical solution spectrum. We provide, within the work presented here, an alternative approach which, although it does not surpass the speed limit for NN systems discussed above, enables significantly faster PST and PSEM than previously reported results \cite{Christandl_2017}. Furthermore, our results provide a range of solutions depending on the NN-to-NNN coupling ratio, thereby broadening the potential pathways towards experimental implementation. 
\label{NNN}
\section{Scaling of power-law long-range quantum speed limits}
\label{scale}
n discussing the speed advantages of these power-law LR PSEM solutions, figure~\ref{fig:4} presents a clear linear scaling for the $\alpha =~3,$ and 6 solutions as a function of total site number $N.$ A non-negligible speed-up is achieved with the dipole–dipole interaction when compared against the speed limit of NN systems \cite{Yung_2006}, whereas the Coulomb interaction provides a remarkable advantage in a sub-linear scaling, $t_m \sim N^{\frac{4}{5}}$. Observing Figure~\ref{fig:4} it can be seen that the time for PSEM in the long-range $\alpha = 1$ model for $N=100$ sites remains below that of a $N=40$ chain for a higher value of $\alpha$. Consequently, this regime allows for state transfer over more than twice the physical distance without incurring a time penalty relative to shorter-range or nearest-neighbour channels. Clearly part of the trade-off lies in the additional control required; we will show in section~\ref{EI} that these requirements are comfortably achievable within state-of-the-art platforms. The QSL of PST, in terms of the maximal coupling of the system, in NN settings is well-known \cite{Yung_2006,Kay_2022}. Here, we discuss the QSL of PSEM in power-law LR models, considering both the numerical scaling of the solutions presented in the preceding sections and the analytical asymptotic behaviour. From the symmetric-spectral argument discussed, we will find correspondence with our numerical findings (figure~\ref{fig_4}) that, counter-intuitively, NN and NNN systems abide by the same QSL.  
\label{scale}
\subsection{Numerical Scaling}
\begin{figure}
    \centering
    \includegraphics[width=0.7\linewidth]{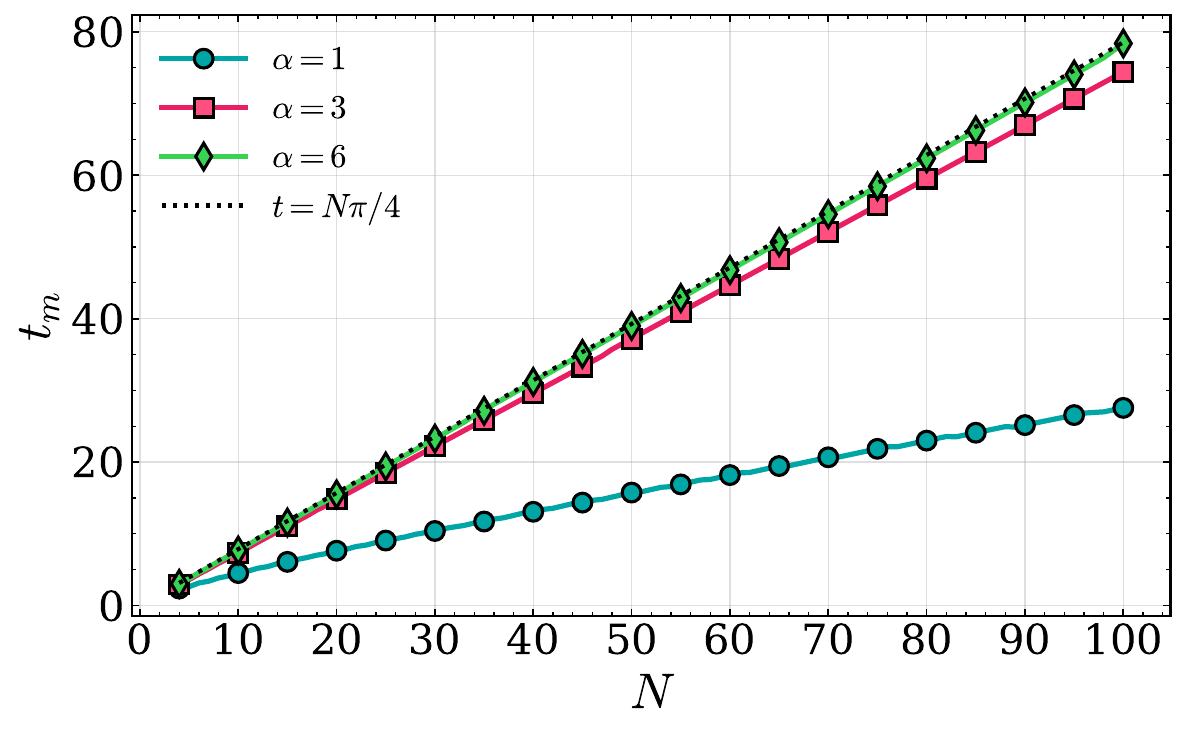}
    \includegraphics[width=0.7\linewidth]{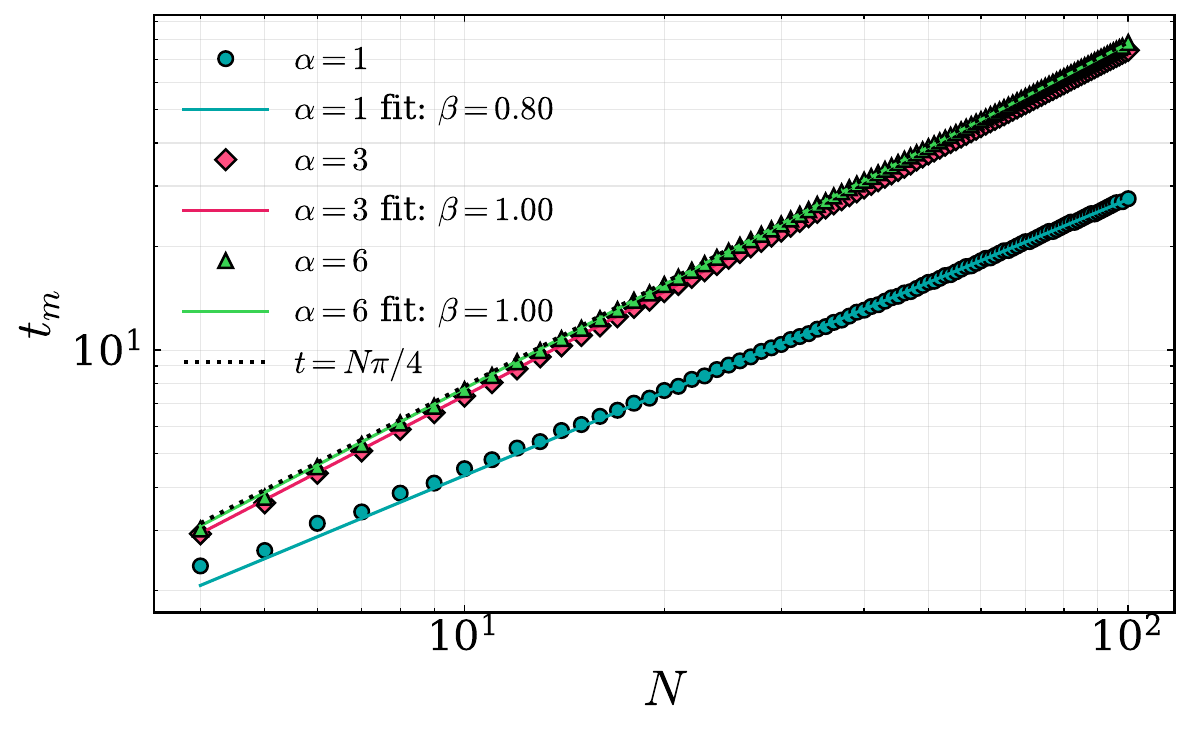}
    \caption{Mirroring time $t_m$ (in units of $J^{-1}_{\mathrm{max}}$) as a function of chain length $N$ for the distinct $\alpha$-decay values considered. The dotted line denotes the speed-limit time $N\pi/4$ for even-site NN chains \cite{Yung_2006}. The corresponding log-log plot (bottom panel) demonstrates the scaling of PAST of power-law LR models within this range of $N$. The mirroring time is fitted to the power-law form $t_m=xN^\beta$, where $\beta$ is the scaling exponent describing how the characteristic time grows with system size. For $\alpha=1$, the fit gives $t_m=0.6816N^{0.802\pm0.002}$ ($R^2=0.9998$), while for $\alpha=3$ and $\alpha=6$, the fits give $t_m=0.7297N^{1.0044\pm0.0003}$ and $t_m=0.7684N^{1.0033\pm0.0004}$, respectively, with $R^2=1.0000$ in both cases.}
    \label{fig:4}
\end{figure}

Here we present the scaling of the results we report in the previous section for power-law decay of $\alpha =1,3$ and $6.$ The data fits remarkably well ($R^2\approx1$) for all linear log-log fits. In figure~\ref{fig:4} we can see the reason behind the slight shifts away from the speed limit for NN chains for $\alpha=3,6$ is due to a difference in the prefactor $b$ (which is $\pi/4$ for NN interactions), whilst for $\alpha = 1$ there is a difference in the prefactor along with a sublinear scaling of $t_m \sim N^{4/5}$, within this range of $N=4-100.$
\subsection{Large-$N$ Scaling $\alpha$}
\begin{figure}[H]
    \centering
    \includegraphics[width=0.8\linewidth]{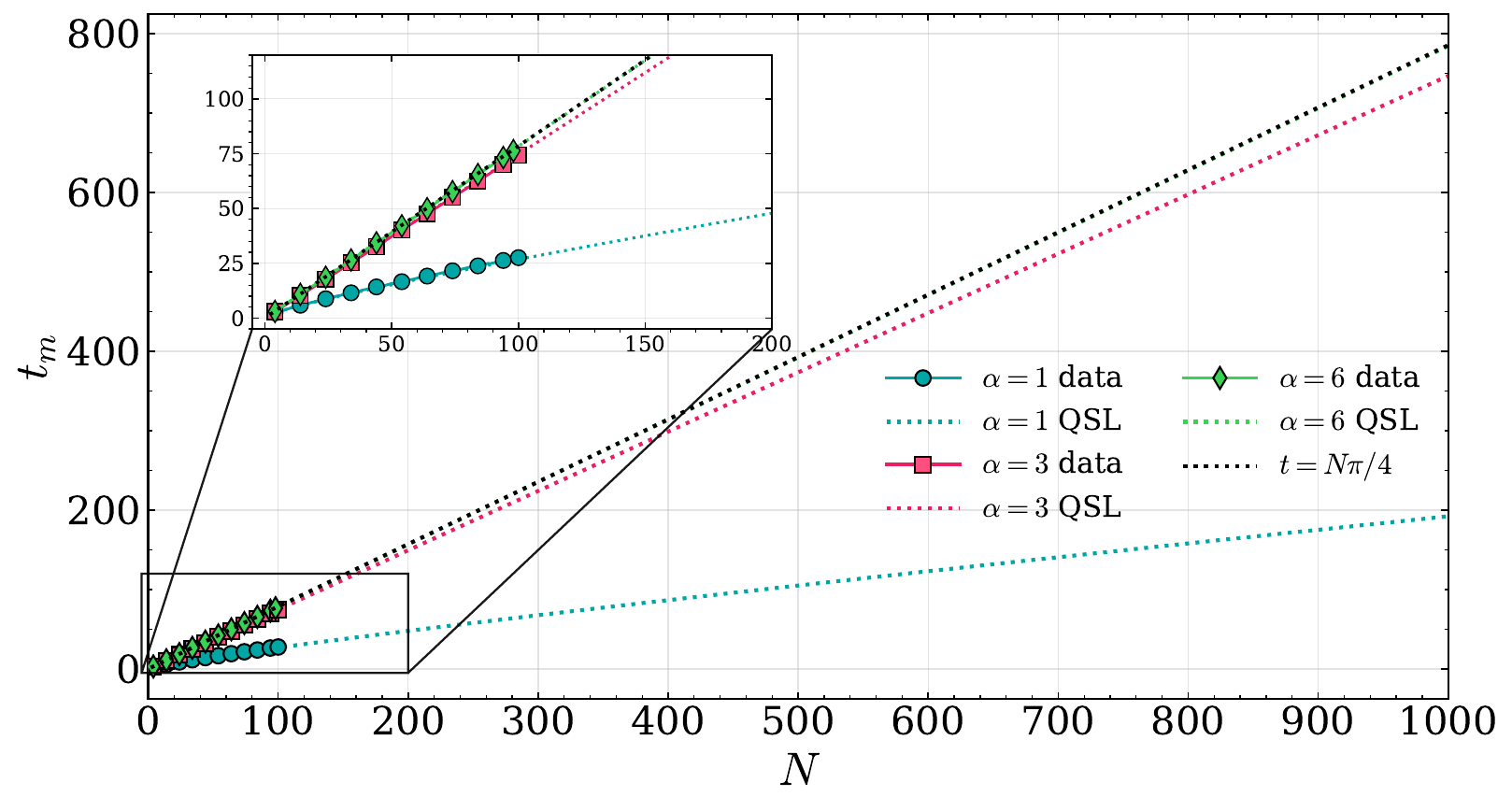}
    \caption{The scaling of the numerical mirroring times from the data of the PSEM-solutions, superimposed on the respective finite-sum QSL (dotted line, equation~\ref{QSL}) of the respective $\alpha-$value for the power-law LR model. We find only minor differences in the prefactor of the numerical scaling and the analytical (for $\alpha = 3,6$ the asymptotic values are 0.746 and 0.784 respectively).}
    \label{fig:scaling}
\end{figure}

Within this subsection, we analytically derive the scaling of the quantum speed limit for PSEM ($t_m$) in the thermodynamic limit, $N\rightarrow\infty$, for the mirror-symmetric power-law LR model. Following the symmetric-spectral argument outlined in \cite{Yung_2006, Kay_2022}, we extend the approach to the power-law LR Hamiltonian and obtain a speed-limit bound in terms of the decay exponent $\alpha$ and chain length $N$. In the large-$\alpha$ limit, the result recovers the known quantum speed limit for the effective NN model. The trace of the product of the mirror operator $M$ and the mirror-symmetric Hamiltonian $H$ can be evaluated in the site basis as
\begin{equation}
\operatorname{Tr}(MH)
=
2\sum_{n=1}^{N/2}J_{N+1-2n}
=
2J_0
\sum_{m=0}^{N/2-1}
\frac{1}{(2m+1)^\alpha}.
\label{trace1}
\end{equation}
As $[H,M]=0$, the eigenstates of $H$ have definite mirror parity and therefore \cite{Kay_2022}
\begin{equation}
\operatorname{Tr}(MH)
=
\sum_{n=1}^{N/2}
\left(
\lambda_{2n-1}-\lambda_{2n}
\right).
\label{trace2}
\end{equation}
Combining equations~\ref{trace1} and \ref{trace2} gives
\begin{equation}
2J_0
\sum_{m=0}^{N/2-1}
\frac{1}{(2m+1)^\alpha}
=
\sum_{n=1}^{N/2}
\left(
\lambda_{2n-1}-\lambda_{2n}
\right).
\label{combined}
\end{equation}
For PSEM, the eigenvalue phases must alternate between the two mirror-parity sectors, requiring $\left(\lambda_{2n-1}-\lambda_{2n}\right)t_m\geq\pi $
for each $n$ (equation~\ref{eig}). Hence,
\begin{equation}
\sum_{n=1}^{N/2}
\left(
\lambda_{2n-1}-\lambda_{2n}
\right)
\geq
\frac{N\pi}{2t_m}.
\label{sum}
\end{equation}
Then simply plugging the inequality~\ref{sum} into equation~\ref{combined} yields the finite-$N$ speed-limit bound
\begin{equation}
t_m \cdot J_{\mathrm{max}}
\geq
\frac{\pi N}
{4\sum_{m=0}^{N/2-1}(2m+1)^{-\alpha}}\label{QSL}
,
\end{equation}
where we have used that for power laws $J_0 = J_{\mathrm{max}}.$
We find very good correspondence with the numerical data from the obtained solutions and the finite-sum QSL of equation~\ref{QSL} (see figure~\ref{fig:scaling}).
Clearly, for sufficiently large \(N\) and $\alpha >1$, the finite sum approaches the corresponding sum over all positive odd integers,
\begin{equation}
\sum_{m=0}^{N/2-1}\frac{1}{(2m+1)^\alpha}
\longrightarrow
\sum_{m=0}^{\infty}\frac{1}{(2m+1)^\alpha},
\qquad N\rightarrow\infty.
\end{equation}
For \(\alpha>1\), the latter sum can be expressed in terms of the Riemann zeta function \cite{Graham_1994} by separating its contributions from the odd and even positive integers,
\begin{equation}
\zeta(\alpha)
=
\sum_{m=0}^{\infty}\frac{1}{(2m+1)^\alpha}
+
\sum_{m=1}^{\infty}\frac{1}{(2m)^\alpha}.
\end{equation}
As $\sum_{m=1}^{\infty}\frac{1}{(2m)^\alpha}
=
2^{-\alpha}\zeta(\alpha),$
we obtain
\begin{equation}
\sum_{m=0}^{\infty}\frac{1}{(2m+1)^\alpha}
=
\left(1-2^{-\alpha}\right)\zeta(\alpha).
\end{equation}
Consequently, in the large-\(N\) limit,
\begin{equation}
\operatorname{Tr}(MH)
\longrightarrow
2J_0\left(1-2^{-\alpha}\right)\zeta(\alpha),
\end{equation}
Substituting this result into the trace-based speed-limit bound gives the asymptotic scaling
\begin{equation}
t_m \cdot J_{\mathrm{max}}
\gtrsim
\frac{\pi N}
{4\left(1-2^{-\alpha}\right)\zeta(\alpha)},
\qquad N\gg1,\quad \alpha>1.
\end{equation}
Consequently, in the large-\(N\) limit,
\begin{equation}
\operatorname{Tr}(MH)
\longrightarrow
2J_0\left(1-2^{-\alpha}\right)\zeta(\alpha),
\end{equation}
Substituting this result into the trace-based speed-limit bound gives the asymptotic scaling
\begin{equation}
t_m \cdot J_{\mathrm{max}}
\gtrsim
\frac{\pi N}
{4\left(1-2^{-\alpha}\right)\zeta(\alpha)},
\qquad N\gg1,\quad \alpha>1.
\end{equation}
In the limit \(\alpha\rightarrow\infty\), $2^{-\alpha}\rightarrow0, \text{ }
\zeta(\alpha)\rightarrow1,$
and therefore
\begin{equation}
t_m \cdot J_{\mathrm{max}}
\gtrsim
\frac{\pi N}{4},
\label{NNQSL}
\end{equation}
recovering the nearest-neighbour result \cite{Yung_2006}. This asymptotic behaviour shows that the QSLs for PSEM in power-law LR models, $\alpha>1$, distinguish themselves from the NN picture by shifts in the prefactor.
The Riemann zeta function diverges at \(\alpha=1\), where it has a simple pole. For this value, the finite-$N$ sum can be derived from writing it as the odd contributions of the harmonic series,
\begin{equation}
\sum_{m=0}^{N/2-1}\frac{1}{2m+1}
=
A_N-\frac{1}{2}A_{N/2},
\end{equation}
where $A_N=\sum_{n=1}^{N}1/n$ denotes the $N^{\mathrm{th}}$ harmonic number \cite{Graham_1994}. Using the asymptotic expansion
\begin{equation}
A_N=\ln N+\gamma+\mathcal{O}(N^{-1}),
\end{equation}
with $\gamma\approx0.5772$ the Euler--Mascheroni constant \cite{finch_2003}, gives
\begin{equation}
\sum_{m=0}^{N/2-1}\frac{1}{2m+1}
=
\frac{1}{2}\ln N+\frac{1}{2}\ln 2+\frac{\gamma}{2}
+\mathcal{O}(N^{-2}).
\end{equation}
Consequently, the finite-$N$ speed-limit bound becomes
\begin{equation}
t_m \cdot J_{\mathrm{max}}
\geq
\frac{\pi N}
{2(\ln N+\ln2+\gamma+\mathcal{O}(N^{-2}))},
\end{equation}
and hence, in the thermodynamic limit,
\begin{equation}
t_m \cdot J_{\mathrm{max}}
\sim
\frac{ N}{\ln N}
. \label{coloumbsl}
\end{equation}
Thus, unlike the $\alpha>1$ cases, for which the odd-power sum approaches a finite value which can lead to a shift in the prefactor alone (as it is simply ``shifted down" by the value of the Riemann zeta function and corresponding $2^{-\alpha}$ term) of the linear-scaling with $N$, the $\alpha=1$ expression diverges sub-linearly with a logarithmic factor. Finally, in discussing the speed limit of PSEM for NNN systems, if one takes an $N\times N$ matrix of the kind of equation~(\ref{13}), it can be found using this symmetric-spectral argument that in fact the NNN coupling does not enter the expression (equation ~\ref{combined}), as it does not appear on the anti-diagonal of $H$. This is consistent with the findings in figure.~\ref{fig_4}, which show no difference in the mirroring times, as it occurs at the derived speed limit in NN-settings, despite significant variation in the NN-to-NNN ratio. This provides supporting evidence for a counter-intuitive result that NN and NNN systems abide by the same QSL of PSEM, provided spectral equivalence, despite the higher-connective topological landscape. 
\section{Robustness and Experimental Implementations}
\label{EI}
\subsection{Robustness}
\begin{figure}[H]
    \centering
    \includegraphics[width=0.85\linewidth]{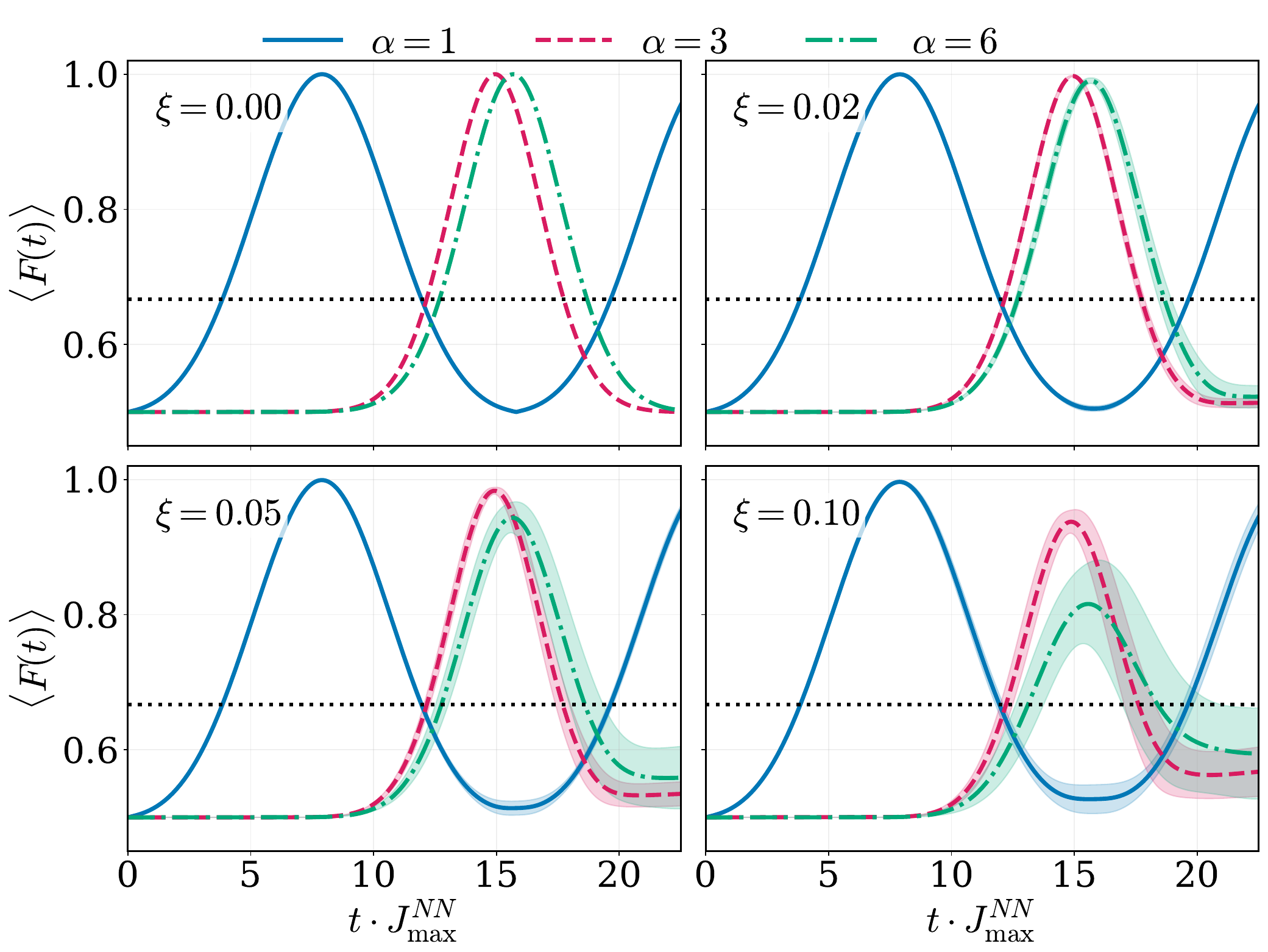}
    \caption{Disorder robustness of the optimised long-range state-transfer chains.
The mean of the average fidelity $\langle F(t)\rangle$ (equation~\ref{AvF}) is shown as a function of the dimensionless time $t\cdot J_{\max}^{NN}$ for $N=20$ and power-law exponents $\alpha=1,3,6$. Each panel corresponds to a different disorder strength $\xi$. For finite $\alpha$, disorder is introduced through independent fluctuations of the on-site energies and inter-site spacings, with $a_i,b_i\sim \mathcal{U}[-0.5,0.5]$. The shaded regions indicate $\pm1\sigma$ over the $R=1000$ disorder realisations. The horizontal dotted line denotes the classical communications fidelity threshold $F=2/3$ \cite{Horodecki}.}
    \label{fig8}
\end{figure}
To assess the experimental feasibility, we have investigated the effect of random asymmetric disorder (i.e., disorder that does not preserve mirror symmetry) on both the position profile and on-site energies of the PSEM LR chain solutions. The on-site energies are modified as
\begin{equation}
B_i^{\mathrm{eff}} = B_i + \xi \,a_i, \quad i = 1, \ldots, N,
\end{equation}
\begin{equation*}
a_i \sim \mathcal{U}\left[-0.5, 0.5\right],
\end{equation*}
where $\xi$ denotes the disorder strength (e.g., $\xi = 0.1$ corresponds to a 10\% fluctuation amplitude). Similarly, the inter-site distances are perturbed according to
\begin{equation}
r_{i}^{\mathrm{eff}} = r_{i} \left(1 + \xi b_i \right), \quad i = 1, \ldots, N-1,
\end{equation}
\begin{equation*}
b_i \sim \mathcal{U}\left[-0.5, 0.5\right],
\end{equation*}
where $a_i$ and $b_i$ are independent random variables and $\mathcal{U}$ is the uniform sampling within the range specified. For the results presented in figure \ref{fig8}, the randomly perturbed effective on-site energies and positions are averaged over 1000 realisations ($R$), to collect the mean fidelity values and standard error. Consistent with the minimalist engineering protocols within \cite{Marshall_2026}, we find in general the higher-$\alpha$ solutions exhibit greater sensitivity to random disorder. However, we note even when the disorder strength $\xi$ allows for $10\%$ error in both the positions and on-site energies, that the lowest mean average fidelity is still comfortably above the LOCC limit of $\frac{2}{3}$ \cite{Horodecki}.   
\subsection{Experimental Realisation}
\begin{figure}[H]
    \centering
    \includegraphics[width=0.8\linewidth]{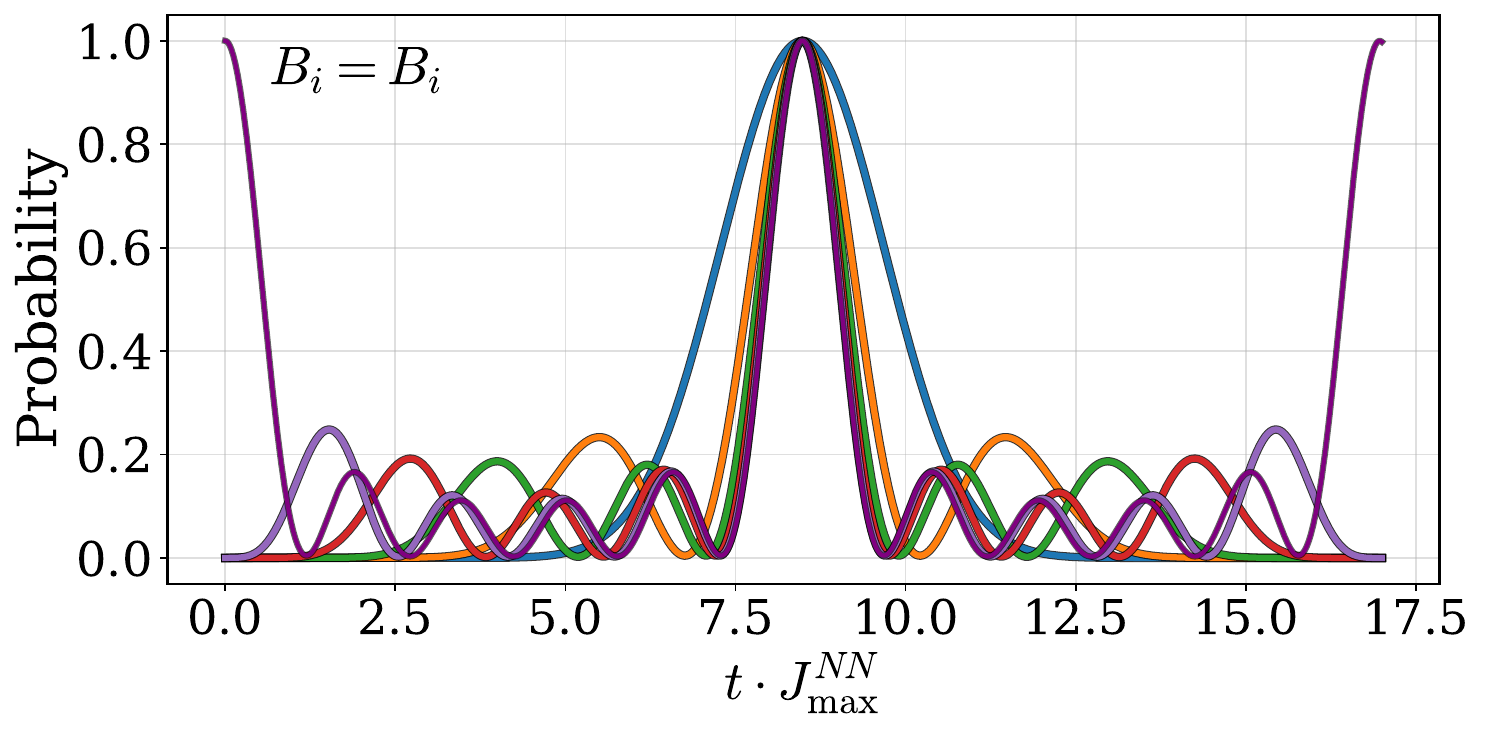}\\[-14mm]
\includegraphics[width=0.8\linewidth]{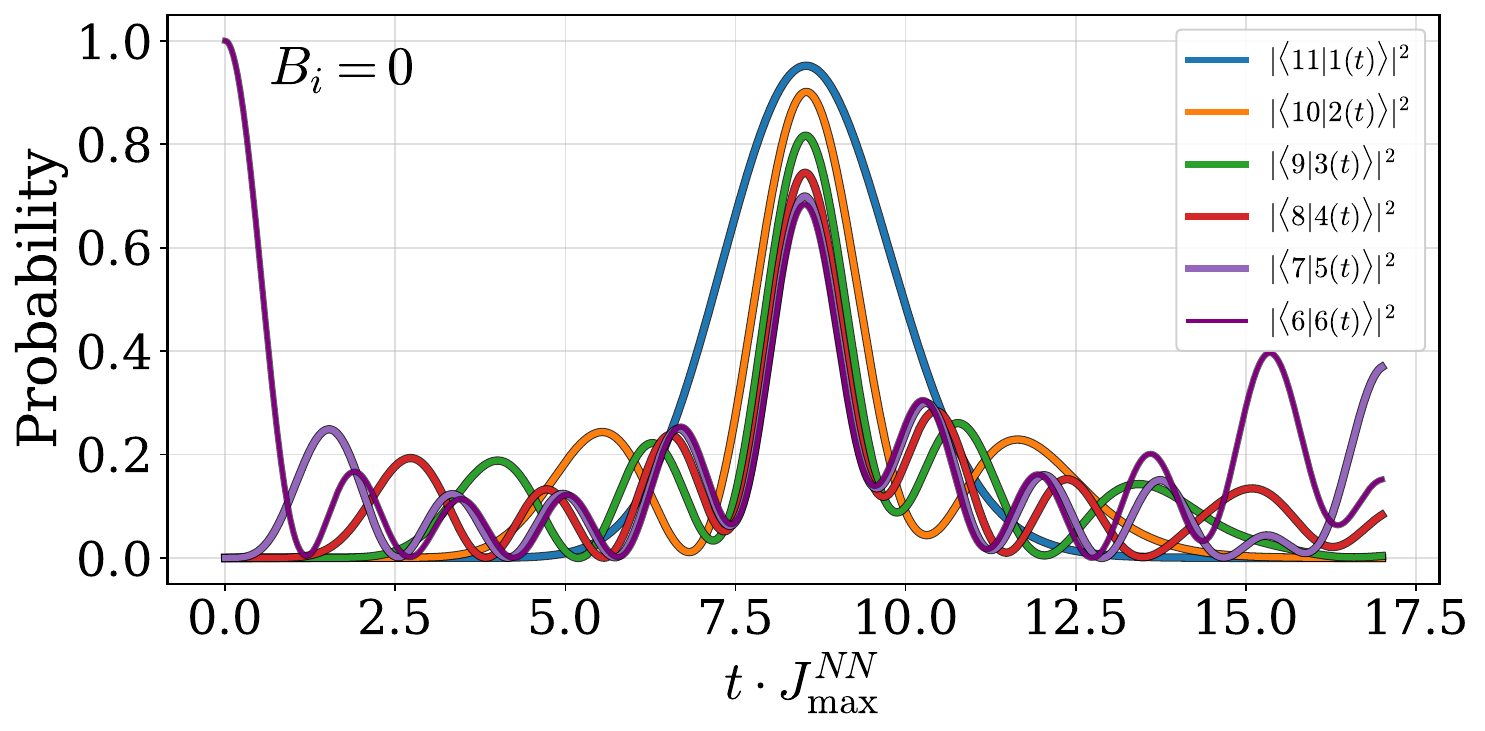}
    \caption{An $\alpha=4$, $N=11$-site power-law LR chain, inspired by the experimental `PST' experiment conducted within \cite{chapman2016}. The upper-panel shows the PSEM allowed by the modulation of both the positions and on-site energies, whilst the lower-panel presents results when only the positions are adjusted within this effective-$\alpha$ LR model. In the latter, we note that the highest peaks remain aligned showing robustness to the $B_i=0$ perturbation.
    (See also figure~\ref{fig_10} in the Appendix).}
    \label{fig_8}
\end{figure}
We may now make our prescriptions more concrete by invoking a specific example of a platform which has been able to generate quantum-advantaged communication using entangled photons and eleven intentionally-spaced waveguides within an array \cite{chapman2016}. The goal of this specific experiment, was to have an effective XX Hamiltonian of only nearest-neighbour interactions and to assume the PSEM coupling (equation~(\ref{Coupling})) within this system. However, the authors note in their discussions that in fact non-negligible higher-order coupling terms participate in the dynamics due to the distant-dependent decay of the couplings. We approximate an effective $\alpha$ value of 4 (See Appendix \ref{effalph}) in order to map this platform to our power-law LR model. The corresponding positions and on-site energies (from the edges) up to four significant figures are as follows:\[
\begin{array}{|c|c|c|c|c|c|}
\scriptsize
 i& 1 & 2 & 3 & 4 & 5 \\ \hline
r_i & 1.330 & 1.240 & 1.198 & 1.176 & 1.166 \\ \hline
B_i(J^{\mathrm{NN}^{-1}}_{\max}) & 0.3631 & 0.2062 & 0.1111 & 0.04867 & 0.01219
\end{array}
\]for PSEM within this system. Figure~\ref{fig_8} shows that for this chain PSEM is possible when one is able to tune the on-site energies accordingly to maintain the linearity of the spectrum, and if one cannot do this we attain the lower-panel which shows diminished transfer and mirroring capacity across the whole chain. We note a close agreement between what is observed numerically and experimentally within \cite{chapman2016}. Taking the positions and converting to real physical distances used within this experiment, and taking $d_{min}$ in their notation equal to the minimum spacing $r_5$, we find the spacings should be: $r_i (\mu \text{m}) =13.69,\ 12.76,\ 12.33,\ 12.10,\ 12.00$. Indeed, this requires only slight spacing differences with what was done already experimentally ($\pm 1.2 \mu\text{m}$ at the edge spacings). The required on-site energy shifts simply follow the profile listed above. Given that current state-of-the-art electro-optic tuning capabilities achieve relative control bounds up to $\sim 10^1 \cdot B_i / J^{\mathrm{NN}}_{\max}$ \cite{Yang_2024}, our prescribed maximum profile of $\sim 10^{-1} \cdot B_i / J^{\mathrm{NN}}_{\max}$ sits comfortably two orders of magnitude below the experimental threshold and poses no significant engineering bottleneck. This provides a reasonable prospect for experimentally witnessing PSEM/total state mirroring via this platform as the ability to tune the `on-site energies' of waveguides can be achieved through either thermo-optic phase shifting \cite{Parra_2024} and/or potential differences introduced by coupled electrodes \cite{Yang_2024}. 

The focus on positioning of individual particles within optical lattice potentials and/or via optical tweezer placements allows for additional flexibility which has been lacking in requiring active suppression of beyond NN inter-qubit strengths. The harmonic-like on-site energy profiles along the system sizes observed $N\sim10-10^2$ as a function of decay constant, for both power-law LR and NNN models, asymptotically approaches $\Delta B_{max}\rightarrow10\cdot J_{max}$ as $\alpha \rightarrow1.$ Within ionic and atomic hardware, the use of harmonic/anharmonic potentials are commonplace \cite{Johanning_2016,Weitenberg2011,Nagies2025}. Directly, the maximal on-site energy strengths shown, in terms of the maximal coupling, is achievable \cite{Nagies2025}. 
In fact, within ionic hardware \cite{Johanning_2016,Welzel_2011} dependent upon the tuning of the magnetic field gradients, the ions spin-spin couplings naturally take the form $J_{ij} \propto \frac{g_J^2 \mu_B^2}{\hbar \, \omega_z^2} \, b^2 \, (A^{-1})_{ij}$ where
$B_i \sim \mu_B g_J b_0$, which leads to a spin-spin coupling to local magnetic field ratio which can reach as high as $\Delta B_i/J_{i,j} \sim 10^4$ (See Appendix \ref{ion}). Within atomic setups, controlled offsets of the optical lattice potentials can be used to tune site-specific qubit energies, with on-site energy shifts of approximately \(B_i/J_{\max}\sim10^{-1}-10^0\) being experimentally accessible \cite{Weitenberg2011,Bloch_2013,Weidner_2025}. Accounting for the observed \(N\)-dependent trends in the required fields, we find that, for the system sizes considered here, this range is sufficient to realise PSEM. In particular, for long-range interactions with \(\alpha=6.0\), the maximum site-dependent field differences required for total state mirroring up to \(N=100\) sites remain within this experimentally-accessible energy scale.

Lastly, with respect to the NNN PSEM models, the on-site energies may be tuned in superconducting circuits \cite{Xiang2023,Yan2019Science} via local detunings of the qubit frequencies in an array/2-D architecture. We note the ratio of the on-site profiles-to-maximal-coupling required for PSEM within these effective Hamiltonians falls conclusively and comfortably within the bounds currently used within quantum hardware, well within the same order-of-magnitude of the maximal coupling. Furthermore, tunable couplings in superconducting architectures has already been demonstrated to benchmark quantum state transfer protocols \cite{Li2018,Xiang2024}. Our NNN findings explicitly show that integrating tunable couplings with reasonable local frequency detuning would enable full quantum state mirroring, at the quantum speed limit \cite{Yung_2006}, across scalable superconducting quantum processors.

\section{Conclusions and Outlook}
In conclusion, we have presented distinct long-range models of technological interest which vary in the power-law exponent $\alpha$ and/or NN-to-NNN ratio and enable perfect single-excitation state mirroring and/or perfect state transfer which either equal or exceed the speed limit for quantum communications due to nearest-neighbour interactions alone. This provides a more realistic framework which is not constrained to the nearest-neighbour-coupling-only scenario and which allows for realistic encoding and retrieval and/or storage of quantum information in linear registers made from hardware composed of atoms, ions, superconductors and/or waveguide arrays. We show that, even when the system is approaching a nearly nearest-neighbour hopping model ($\alpha=6$), non-negligible on-site energy profiles (e.g. implemented by local magnetic fields) are required in order to witness perfect single-excitation mirroring. Furthermore, we show lower-$\alpha$ decay constants present speed and robustness advantages at the cost of higher-engineering demands. From a resource perspective, the sub-linear growth in the mirroring time of linear-spectrum Coulomb-interacting LR models, with respect to the system size, highlights a significant operational speed-up. In general, lower-alpha solutions require greater variation in the on-site energy and spacing profiles, but offer increased robustness to disorder in the couplings and on-site energies. This presents a trade-off between greater engineering demands and more robust, significantly faster quantum information transfer, which could help circumvent system size limitations within real hardware.

Current and future work will be to extend the methodology discussed here to explore exotic spectra which may have specific engineering advantages over the linear spectrum \cite{Vinet_2012,Nelmes_2026}. Furthermore, it has been established for quite some time that, via inverse eigenvalue methods, in purely tridiagonal (NN) mirror-symmetric matrices that a numerically stable reconstruction algorithm yields a \textit{unique} solution of the final matrix corresponding to specifically-chosen eigenvalues \cite{Hald_1976,deBoorGolub_1978,Hochstadt_1979}. Though we do not claim that the solutions presented here are `unique' to the spectrum and rank of matrix, as beyond the tridiagonal case, to our knowledge, a uniqueness theorem does not exist. We note, however, for a given $N$ and $\alpha-$value we would find repeating solutions using the iterative eigenvalue solver within this study, up to $10^{-8}$ tolerance. The roles of temperature for the transfer within/of thermalised systems/thermal states are still unclear even within nearest-neighbour systems (barring only a couple of works \cite{Karbach_2005,PhysRevA.71.042330}) let alone longer-range interactions. The transfer of multiple-excitations, also,  remains an area of great interest and current discussion. Work is currently being undertaken to understand how this alternative for PSEM may also be used for the purposes of energy transport \cite{Murphy_2025,krisnanda_2026}, non-stabiliser (`Magic') resource channels \cite{Kitaev_2005,Palamara_2026}, and in relation to other emergent technologies such as quantum-advantaged timekeeping devices \cite{nelmes2_2026}.
\label{CO}

\section*{Acknowledgments}
CCN acknowledges support from EPSRC, grant number is EP\textbackslash W524657\textbackslash1 as well as helpful discussions with Tony John George Apollaro.
\label{ack}
\appendix
\section*{Appendix}
\section{Reduction in Transfer Probability for $B_i=0$}
\label{reduction}
For larger system sizes $N=50-$sites NNN chains of the kind discussed within Sec.~\ref{NNN} for $\alpha=6.3$ where the required magnetic field profile to maintain linearity of the spectrum is omitted. 
\begin{figure}[H]
    \centering
    \includegraphics[width=0.8\linewidth]{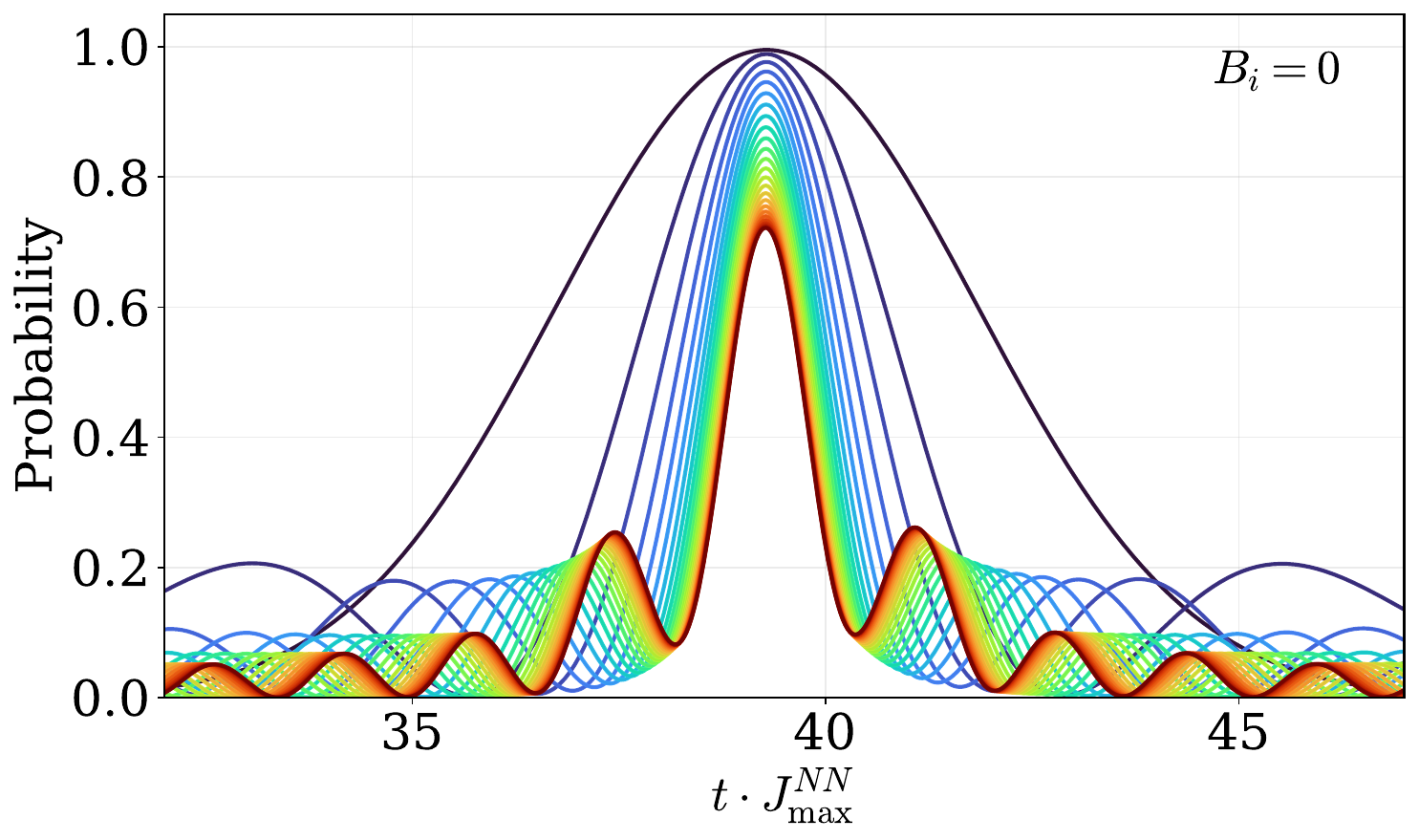}
    \caption{Sites and mirrored sites with significantly reduced probability of transferring information between them once the magnetic field profile is omitted, despite the strengths of the on-site profiles being $\text{max}(\Delta B)=10^{-2}$ in units of $J^{-1}_{max}$.}
    \label{fig_10}
\end{figure}
\begin{table}[H]

\centering
\caption{Maximum transfer fidelities for mirror-symmetric site pairs in an $N=50$; $\alpha=6.3$-chain, when the local magnetic fields/on-site energies are simply `switched off.'}

\setlength{\tabcolsep}{50pt} 

\begin{tabular}{c |c}
\hline
Prob($i \rightarrow j$) =$|\langle j|i(t_m)\rangle|^2$ & Value \\
\hline \hline
$\langle 50|1(t_m)\rangle^2$  & 0.995373 \\
\hline
$\langle 49|2(t_m)\rangle^2$  & 0.988922 \\
\hline
$\langle 48|3(t_m)\rangle^2$  & 0.976757 \\
\hline
$\langle 47|4(t_m)\rangle^2$  & 0.962114 \\
\hline
$\langle 46|5(t_m)\rangle^2$  & 0.945925 \\
\hline
$\langle 45|6(t_m)\rangle^2$  & 0.928854 \\
\hline
$\langle 44|7(t_m)\rangle^2$  & 0.911378 \\
\hline
$\langle 43|8(t_m)\rangle^2$  & 0.893853 \\
\hline
$\langle 42|9(t_m)\rangle^2$  & 0.876557 \\
\hline
$\langle 41|10(t_m)\rangle^2$ & 0.859706 \\
\hline
$\langle 40|11(t_m)\rangle^2$ & 0.843468 \\
\hline
$\langle 39|12(t_m)\rangle^2$ & 0.827978 \\
\hline
$\langle 38|13(t_m)\rangle^2$ & 0.813343 \\
\hline
$\langle 37|14(t_m)\rangle^2$ & 0.799645 \\
\hline
$\langle 36|15(t_m)\rangle^2$ & 0.786949 \\
\hline
$\langle 35|16(t_m)\rangle^2$ & 0.775306 \\
\hline
$\langle 34|17(t_m)\rangle^2$ & 0.764754 \\
\hline
$\langle 33|18(t_m)\rangle^2$ & 0.755325 \\
\hline
$\langle 32|19(t_m)\rangle^2$ & 0.747040 \\
\hline
$\langle 31|20(t_m)\rangle^2$ & 0.739915 \\
\hline
$\langle 30|21(t_m)\rangle^2$ & 0.733964 \\
\hline
$\langle 29|22(t_m)\rangle^2$ & 0.729194 \\
\hline
$\langle 28|23(t_m)\rangle^2$ & 0.725613 \\
\hline
$\langle 27|24(t_m)\rangle^2$ & 0.723223 \\
\hline
$\langle 26|25(t_m)\rangle^2$ & 0.722027 \\
\hline
\end{tabular}
\label{tab3}
\end{table}
\[\]
\section{`Effective power-law exponent' adapted from the coupling reported in \cite{chapman2016}}
\label{effalph}
In seeking to map a more complex, real system to a discrete spin model representation, for the purposes of simulating the dynamics as a power-law LR model, we take the coupling coefficients reported within \cite{chapman2016} and approximate (based on a NN-to-NNN coupling ratio) what power-law long range decay exponent produces similar differences in the couplings. This allows us to understand better how on-site energies may be tuned to transition from QPST-to-PSEM. We can calibrate our long-range model to provide an approximate description of any experimental system for which the couplings beyond NN are dominated by the NNN terms, with the higher terms much smaller. We do this by first choosing an effective $\alpha$ that produces the correct ratio between NN and NNN couplings. The actual energy scale for both is then set by choosing a suitable $J_0$, which in turn then provides the energy scale for any site-specific magnetic fields/on-site energies $B_i$ that might be required. For the experimental parameters $N=11$ and $d_{\min}=12\,\mu\mathrm{m}$,
the waveguide separations are given by
\begin{equation}
d_n=d_{\min}+\frac{1}{b}
\ln\left[\frac{1}{2}\frac{n(N-n)}{N^2-1}\right],
\end{equation}
with $b=0.19\,\mu\mathrm{m}^{-1}$. This gives
\begin{equation}
d_{1,2}\simeq14.891\,\mu\mathrm{m},\qquad
d_{2,3}\simeq13.344\,\mu\mathrm{m}.
\end{equation} 
Hence the nearest-neighbour and next-nearest-neighbour distances are
\begin{equation}
r_{\mathrm{NN}} \equiv r_{1}=d_{1,2}\simeq14.891\,\mu\mathrm{m},
\end{equation}
and
\begin{equation}
r_{\mathrm{NNN}} \equiv (r_{1} + r_{2})=(d_{1,2}+d_{2,3})\simeq28.235\,\mu\mathrm{m}.
\end{equation}
In fact, the experimental coupling scales as (See Supplementary Materials of \cite{chapman2016})
\begin{equation}
J(r)\propto e^{-br}.
\end{equation}
To approximate this by $
J(r)\propto r^{-\alpha},$
we match the NN-to-NNN coupling ratio, giving
\begin{equation}
e^{b(r_{\mathrm{NNN}}-r_{\mathrm{NN}})}
=
\left(\frac{r_{\mathrm{NNN}}}{r_{\mathrm{NN}}}\right)^{\alpha_{\mathrm{eff}}},
\end{equation}
so that
\begin{equation}
\alpha_{\mathrm{eff}}
=
\frac{b(r_{\mathrm{NNN}}-r_{\mathrm{NN}})}
{\ln(r_{\mathrm{NNN}}/r_{\mathrm{NN}})}.
\end{equation}
Substituting the above values gives $
\alpha_{\mathrm{eff}}
=
\frac{0.19(28.235-14.891)}
{\ln(28.235/14.891)}
\simeq3.96 \approx4.$ Therefore, to approximate the
$N=11$, $d_{\min}=12\,\mu\mathrm{m}$ experimental array using our long-range model, it is reasonable
to approximate the model coupling by
\begin{equation}
J_{ij}=\frac{J_0}{d_{ij}^{4}}.
\end{equation}
We stress that this mapping is approximate and is based specifically on matching an NN-to-NNN coupling
ratio, rather than implying that the exponential and power-law forms are
identical at all distances.

\section{Ionic typical $B_i/J_{i,j}$-values}
\[\]
\label{ion}
We consider spin-spin interactions of the form
\begin{equation}
J_{ij} \propto
\frac{g_J^2\mu_B^2}{m\omega_z^2},
b'^2,(A^{-1})_{ij},
\end{equation}
together with site-dependent magnetic fields
\begin{equation}
B_i \simeq g_J\mu_B b(z_i),
\end{equation}
as realized in ion-trap architectures employing magnetic-field gradients \cite{Johanning_2016,Welzel_2011, Nagies2025}. Here, \(J_{ij}\) and \(B_i\) are understood as energy scales. The relevant experimental parameters are of order
\begin{equation}
g_J\sim\mathcal{O}(1),\qquad
\frac{\mu_B}{h}\simeq1.4\times10^{10}\ {\rm Hz/T},
\end{equation}
\begin{equation}
b'\sim\mathcal{O}(10)\ {\rm T/m},\qquad
\omega_z\sim\mathcal{O}(10^{5})\ {\rm Hz},
\end{equation}
leading to spin-spin couplings of order
\begin{equation}
\frac{J}{h}\sim\mathcal{O}(10-10^2)\ {\rm Hz}.
\end{equation}
For a magnetic field varying approximately linearly along the ion chain,
\begin{equation}
b(z)=b_0+b'z,
\end{equation}
the absolute Zeeman energy and the site-to-site energy difference play distinct roles. The former is
\begin{equation}
B_i\simeq g_J\mu_B b_0,
\end{equation}
whereas for neighbouring ions separated by a distance \(a\),
\begin{equation}
\Delta B\equiv |B_{i+1}-B_i|
\simeq g_J\mu_B b'a.
\end{equation}
Thus, it is \(\Delta B/J\), rather than \(B/J\), that determines the relative strength of the spatially varying on-site energies in the spin-chain model. For representative ion separations \(a\sim1~\mu{\rm m}\) and magnetic-field gradients \(b'\sim10~{\rm T/m}\), the gradient-induced Zeeman-energy difference is of order
\begin{equation}
\frac{\Delta B}{h}
\sim
g_J\frac{\mu_B}{h}b'a
\sim\mathcal{O}(10^5)\ {\rm Hz},
\end{equation}
which is substantially larger than the characteristic spin-spin coupling \(J/h\sim10-10^2~{\rm Hz}\). Consequently, a regime with
\begin{equation}
\frac{\Delta B}{J}\sim10
\end{equation}
is well within the energy scales accessible in magnetic-gradient ion-trap systems, given appropriately-chosen magnetic field gradients. We therefore argue \(\Delta B/J=10\) is a realistic parameter regime for the site-dependent fields in our simulations.

\bibliography{bib}

\end{document}